\documentclass{scrartcl}

\usepackage{amsmath,amssymb}
\usepackage{fullpage}
\usepackage{enumitem}
\usepackage[hidelinks]{hyperref}
\usepackage{pifont}

\usepackage[xindy]{glossaries}	
\setkeys{glslink}{hyper=false}
\newacronym{fcfs}{FCFS}{first-come, first-served}
\newacronym{fifo}{FIFO}{first-in, first-out}
\newacronym{ps}{PS}{processor sharing}
\newacronym{oi}{OI}{order-independent}
\newacronym{jiq}{JIQ}{Join-Idle-Queue}
\newacronym{alis}{ALIS}{\textit{assign longest idle server}}

\def\arrivals{{\Sigma}}
\def\departures{{\Gamma}}
\def\classes{{\mathcal{I}}}
\def\types{{\mathcal{K}}}
\def\servers{{\mathcal{S}}}
\def\seq{{\mathcal{C}}}
\def\states{{\mathcal{X}}}
\def\A{{\mathcal{A}}}

\def\N{{\mathbb{N}}}
\def\R{{\mathbb{R}}}
\def\espace{{\mathcal{S}}}

\usepackage{amsthm}

\theoremstyle{definition}

\theoremstyle{remark}

\usepackage{xcolor}
\definecolor{myred}{RGB}{255,0,0}
\definecolor{myblue}{RGB}{92,92,235}
\definecolor{myorange}{RGB}{253,188,64}
\definecolor{mygreen}{RGB}{154,205,50}
\definecolor{mygray}{RGB}{210,210,210}

\definecolor{myorangeplot}{RGB}{245,138,42}
\definecolor{mygreenplot}{RGB}{0,200,0}

\usepackage{graphics,graphicx,standalone}
\usepackage{subfig} 
\usepackage{tikz,tikz-3dplot,pgfplots}
\usetikzlibrary{calc,arrows,shapes,snakes,backgrounds}

\usepackage{pgfplots,pgfplotstable}
\pgfplotsset{compat=1.8}
\usepgfplotslibrary{fillbetween}

\tikzset{
  class/.style = {draw, rounded corners=.05cm, minimum size=.65cm},
  type/.style = {draw, rounded corners=.05cm, minimum size=.8cm, regular polygon, regular polygon sides=6},
  smalltype/.style = {draw, rounded corners=.04cm, minimum size=.75cm, regular polygon, regular polygon sides=6},
  server/.style = {draw, circle, minimum size=.7cm},
  serverps/.style = {draw, circle, minimum size=1.1cm},
  serveroi/.style = {draw, circle, minimum size=1.3cm},
  graphnode/.style = {draw, circle, minimum size=.8cm},
  fcfs/.style={
    draw,
    rectangle split, rectangle split parts=#1,
    rectangle split horizontal,
    rectangle split empty part width=-.17cm, rectangle split empty part height=.65cm,
    rounded corners=.05cm,
  },
  ps/.style={
    rotate=90,
    draw,
    rectangle split, rectangle split parts=#1,
    rectangle split horizontal,
    rectangle split empty part width=-.17cm, rectangle split empty part height=.9cm,
    rounded corners=.05cm,
  },
  arrow/.style={
    ->,
    rounded corners=.05cm,
  },
}

\pgfplotsset{
  table/search path={./},
  defaultplots/.style={
    xlabel near ticks,
    ylabel near ticks,
    grid=major,
    xmin=0, xmax=4,
    ymin=0, ymax=1,
    xlabel={Load $\rho$},
    ylabel={Job and server metrics},
    x label style = {yshift=0.15cm},
    xticklabel style = {font=\footnotesize},
    yticklabel style = {font=\footnotesize},
    label style = {font=\footnotesize},
    legend style={
      rounded corners=.05cm,
      cells={anchor=west, align=left},
      font=\footnotesize,
    },
    width=.6\linewidth, height=.35\linewidth,
  },
  every axis plot/.append style={line width=.7pt},
}

\begin{document}

\title{Dynamic Load Balancing with Tokens\thanks{\copyright~IFIP, 2018.
This is the author's version of the work.
It is posted here by permission of IFIP for your personal use.
Not for redistribution.
The definitive version will be published
in IFIP NETWORKING 2018 proceedings.}}
\author{C\'eline Comte \\
Nokia Bell Labs and T\'el\'ecom ParisTech, University Paris-Saclay, France \\
celine.comte@nokia.com
}

\date{\today}
\maketitle

\begin{abstract}
  Efficiently exploiting the resources of data centers
  is a complex task that
  requires efficient and reliable
  load balancing and resource allocation algorithms.
  The former are in charge of assigning jobs to servers upon their arrival in the system,
  while the latter are responsible for sharing server resources between their assigned jobs.
  These algorithms should take account of various constraints, such as data locality, that restrict the feasible job assignments.
  In this paper,
  we propose a token-based mechanism that efficiently balances load between servers
  without requiring any knowledge on job arrival rates and server capacities.
  Assuming a balanced fair sharing of the server resources,
  we show that the
  resulting dynamic load balancing
  is insensitive to the job size distribution.
  Its performance is compared to that obtained
  under the best static load balancing
  and in an ideal system that would constantly optimize the resource utilization.
\end{abstract}

\section{Introduction}
\label{sec:intro}

The success of cloud services encourages operators
to scale out their data centers and optimize the resource utilization.
The current trend consists in virtualizing applications
instead of running them on dedicated physical resources \cite{B13}.
Each server may then process several applications in parallel
and each application may be distributed among several servers.
Better understanding the dynamics
of such server pools
is a prerequisite for developing
load balancing and resource allocation policies
that fully exploit this new degree of flexibility.

Some recent works have tackled this problem
from the point of view of queueing theory \cite{alis,SV15,G15,BC17}.
Their common feature is the adoption of a bipartite graph
that translates practical constraints such as data locality
into compatibility relations between jobs and servers.
These models apply in various systems such as
computer clusters, where the shared resource is the CPU \cite{G15,BC17},
and content delivery networks, where the shared resource is the server upload bandwidth \cite{SV15}.
However,
these pool models do not
consider simultaneously the impact of complex load balancing and resource allocation policies.
The model of \cite{alis} lays emphasis on dynamic load balancing,
assuming neither server multitasking nor job parallelism.
The bipartite graph describes the initial compatibilities of incoming jobs,
each of them being eventually assigned to a single server.
On the other hand,
\cite{G15,SV15,BC17} focus on the problem of resource allocation, assuming
a static load balancing that assigns incoming jobs to classes at random, independently of the system state.
The class of a job in the system identifies the set of servers that can be pooled to process it in parallel.
The corresponding bipartite graph, connecting classes to servers,
restricts the set of feasible resource allocations.

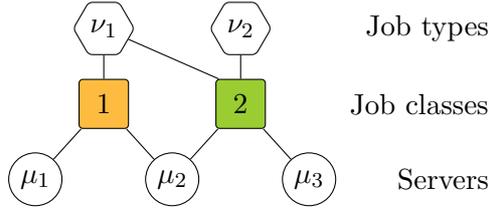
\begin{figure}
  \centering
  \begin{tikzpicture}
    \def\width{1.8cm}
    \def\height{1cm}

    \node[server] (server1) {};
    \node[server] (server2) at ($(server1)+(\width,0)$) {};
    \node[server] (server3) at ($(server2)+(\width,0)$) {};
    \node at (server1) {$\mu_1$};
    \node at (server2) {$\mu_2$};
    \node at (server3) {$\mu_3$};

    \node[class,fill=myorange] (class1) at ($(server1)!.5!(server2)+(0,\height)$) {1};
    \node[class,fill=mygreen] (class2) at ($(server3)!.5!(server2)+(0,\height)$) {2};

    \node[type] (type1) at ($(class1)+(0,\height)$) {};
    \node[type] (type2) at ($(class2)+(0,\height)$) {};
    \node at (type1) {$\nu_1$};
    \node at (type2) {$\nu_2$};

    \draw[-] (server1) -- (class1) -- (server2) -- (class2) -- (server3);
    \draw[-] (class1) -- (type1) -- ($(class2.north)-(.29cm,0)$) (class2) -- (type2);

    \node[anchor=east] at ($(server3)+(2.5cm,0)$) {Servers};
    \node[anchor=east] at ($(server3)+(2.5cm,0)+(0,\height)$) {Job classes};
    \node[anchor=east] at ($(server3)+(2.5cm,0)+(0,2*\height)$) {Job types};
  \end{tikzpicture}
  \caption{
    A compatibility graph between types, classes and servers.
    Two consecutive servers can be pooled to process jobs in parallel.
    Thus there are two classes,
    one for servers $1$ and $2$
    and another for servers $2$ and $3$.
    Type-$1$ jobs can be assigned to any class,
    while type-$2$ jobs can only be assigned to the latter.
    This restriction may result
    from data locality constraints for instance.
  }
  \label{fig:tripartite}
\end{figure}

In this paper,
we introduce a tripartite graph that explicitly differentiates
the compatibilities of an incoming job from its actual assignment by the load balancer.
This new model
allows us to study the joint effect of
load balancing
and resource allocation.
A toy example is shown in \figurename~\ref{fig:tripartite}.
Each incoming job has a type that defines its compatibilities;
these may reflect its parallelization degree or locality constraints, for instance.
Depending on the system state,
the load balancer matches the job with a compatible class
that subsequently determines its assigned servers.
The upper part of our graph, which puts constraints on load balancing,
corresponds to the bipartite graph of \cite{alis};
the lower part, which restricts the resource allocation,
corresponds to the bipartite graph of \cite{G15,SV15,BC17}.

We use this
new framework to study
load balancing and resource allocation policies that are \emph{insensitive},
in the sense that they make the system performance independent of fine-grained traffic characteristics.
This property is highly desirable
as it allows service providers to dimension their infrastructure
based on average traffic predictions only.
It has been extensively studied in the queueing literature \cite{BP02,BP03-1,BJP04,SV15}.
In particular,
insensitive load balancing policies were introduced in \cite{BJP04} in a generic queueing model,
assuming an arbitrary insensitive allocation of the resources.
These load balancing policies were defined as
a generalization of the static load balancing described above,
where the assignment probabilities of jobs to classes
depend on both the job type and the system state,
and are chosen to
preserve insensitivity.

Our main contribution is an algorithm based on tokens
that enforces such an insensitive load balancing
without performing randomized assignments.
More precisely,
this is a \emph{deterministic} implementation
of an insensitive load balancing
that adapts dynamically to the system state,
under an arbitrary compatibility graph.
The principle is as follows.
The assignments are regulated through a bucket containing a fixed number of tokens of each class.
An incoming job seizes the longest available token among those that identify a compatible class,
and is blocked if it does not find any.
The rationale behind this algorithm is to use
the release order of tokens as an information on the relative load of their servers:
a token that has been available for a long time without being seized
is likely to identify a server set that is less loaded than others.
As we will see,
our algorithm mirrors the \gls{fcfs} service discipline proposed in \cite{BC17} to implement balanced fairness,
which was defined in \cite{BP03-1} as the most efficient insensitive resource allocation.

The closest existing algorithm we know is \gls{alis},
introduced in reference \cite{alis} cited above.
This work focuses on server pools without job parallel processing nor server multitasking.
Hence, \gls{alis} can be seen as a special case of our algorithm
where each class identifies a server with a single token.
The algorithm we propose is also related to
the blocking version of Join-Idle-Queue \cite{jiq} studied in \cite{BBL17}.
More precisely, we could easily generalize our algorithm to server pools with several load balancers,
each with their own bucket.
The corresponding queueing model,
still tractable using known results on networks of quasi-reversible queues \cite{kelly},
extends that of \cite{BBL17}.

\paragraph*{Organization of the paper}

Section \ref{sec:resource-allocation} recalls known facts about resource allocation in server pools.
We describe a standard pool model based on a bipartite compatibility graph
and explain how to apply balanced fairness in this model.
Section \ref{sec:load-balancing} contains our main contributions.
We describe our pool model based on a tripartite graph
and introduce a new token-based insensitive load balancing mechanism.
Numerical results are presented in Section \ref{sec:num}.

\section{Resource allocation}
\label{sec:resource-allocation}

We first recall the model considered in \cite{G15,SV15,BC17}
to study the problem of resource allocation in server pools.
This model will be extended in Section \ref{sec:load-balancing}
to integrate dynamic load balancing.

\subsection{Model}
\label{subsec:bipartite}

We consider a pool of $S$ servers.
There are $N$ job classes and we let $\classes = \{1,\ldots,N\}$ denote the set of class indices.
For now, each incoming job is assigned to a compatible class at random, independently of the system state.
For each $i \in \classes$, the resulting arrival process of jobs assigned to class $i$
is assumed to be Poisson with a rate $\lambda_i > 0$ that may depend on
the job arrival rates, compatibilities and assignment probabilities.
The number of jobs of class $i$ in the system is limited by $\ell_i$, for each $i \in \classes$,
so that a new job is blocked if its assigned class is already full.
Job sizes are independent and exponentially distributed with unit mean.
Each job leaves the system immediately after service completion.

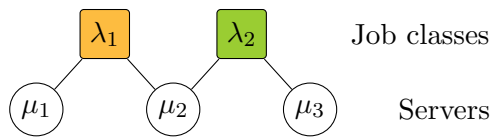
\begin{figure}[h]
  \centering
  \begin{tikzpicture}
    \def\width{1.8cm}
    \def\height{1cm}

    \node[server] (server1) {};
    \node[server] (server2) at ($(server1)+(\width,0)$) {};
    \node[server] (server3) at ($(server2)+(\width,0)$) {};
    \node at (server1) {$\mu_1$};
    \node at (server2) {$\mu_2$};
    \node at (server3) {$\mu_3$};

    \node[class,fill=myorange] (class1) at ($(server1)!.5!(server2)+(0,\height)$) {$\lambda_1$};
    \node[class,fill=mygreen] (class2) at ($(server3)!.5!(server2)+(0,\height)$) {$\lambda_2$};

    \draw[-] (server1) -- (class1) -- (server2) -- (class2) -- (server3);

    \node[anchor=east] at ($(server3)+(2.5cm,0)$) {Servers};
    \node[anchor=east] at ($(server3)+(2.5cm,0)+(0,\height)$) {Job classes};
  \end{tikzpicture}
  \caption{A compatibility graph between classes and servers.
  Servers $1$ and $3$ are dedicated,
  while server $2$ can serve both classes.
  The server sets associated with classes $1$ and $2$ are $\servers_1 = \{1,2\}$ and $\servers_2 = \{2,3\}$, respectively.}
  \label{fig:bipartite}
\end{figure}

The class of a job defines the set of servers that can be pooled to process it.
Specifically, for each $i \in \classes$,
a job of class $i$ can be served in parallel by any subset of servers within the non-empty set $\servers_i \subset \{1,\ldots,S\}$.
This defines a bipartite compatibility graph between classes and servers,
where there is an edge between a class and a server
if the jobs of this class can be processed by this server.
\figurename~\ref{fig:bipartite} shows a toy example.

When a job is in service on several servers, its service rate is the sum of the rates allocated by each server to this job.
For each $s = 1,\ldots,S$,
the capacity of server $s$ is denoted by $\mu_s > 0$.
We can then
define a function $\mu$ on the power set of $\classes$ as follows:
for each $\A \subset \classes$,
$$
\mu(\A) = \sum_{s \in \bigcup_{i \in \A} \servers_i} \mu_s
$$
denotes the aggregate capacity of the servers that can process at least one class in $\A$,
i.e., the maximum rate at which jobs of these classes can be served.
$\mu$ is a submodular, non-decreasing set function \cite{F05}.
It is said to be normalized because $\mu(\emptyset) = 0$.

\subsection{Balanced fairness}
\label{subsec:bf}

We first recall the definition of balanced fairness \cite{BP03-1},
which was initially applied to server pools in \cite{SV15}.
Like \gls{ps} policy, balanced fairness assumes that
the capacity of each server can be divided continuously between its jobs.
It is further assumed that the resource allocation only depends on the number of jobs of each class in the system;
in particular, all jobs of the same class receive service at the same rate.

The system state is described by the vector
$x = (x_i : i \in \classes)$ of numbers of jobs of each class in the system.
The state space is $\states = \{x \in \N^N : x \le \ell\}$,
where $\ell = (\ell_i : i \in \classes)$ is the vector of per-class constraints
and the comparison $\le$ is taken componentwise.
For each $i \in \classes$,
we let $\phi_i(x)$ denote the total service rate allocated to class-$i$ jobs in state $x$.
It is assumed to be nonzero if and only if $x_i > 0$,
in which case each job of class $i$
receives service at rate
$\phi_i(x) / x_i$.

\paragraph*{Queueing model}

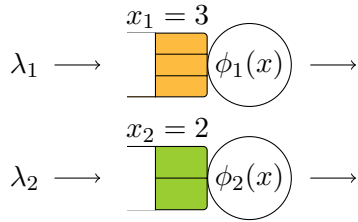
\begin{figure}[h]
  \centering
  \begin{tikzpicture}
    \node (queue1) {};
    \node (queue2) at ($(queue1)+(0,-1.5cm)$) {};

    \node[ps=3,inner ysep=.15cm,rectangle split empty part width=.02cm,fill=myorange] (queue1) at (queue1) {};
    \fill[white] ([xshift=-\pgflinewidth-1pt,yshift=-\pgflinewidth+1pt] queue1.north east)
    rectangle ([xshift=\pgflinewidth+4pt,yshift=\pgflinewidth-1pt] queue1.north west);
    \fill[white] ([xshift=-\pgflinewidth-1pt,yshift=-\pgflinewidth] queue1.north east)
    rectangle ([xshift=15pt,yshift=\pgflinewidth] queue1.north west);
    \draw[-] ([xshift=15pt] queue1.north east) -- ([xshift=15pt] queue1.north west);

    \node[serverps] (phi1) at ($(queue1.south)+(.55cm,0)$) {};
    \node at (phi1) {$\phi_1(x)$};
     
    \node[anchor=south] at ($(queue1)+(.08cm,.35cm)$) {$x_1 = 3$};

    \draw[arrow] ($(queue1.north)+(-.8cm,0)$) -- node[pos=-.1,anchor=east] {$\lambda_1$} ($(queue1.north)-(.2cm,0)$);
    \draw[arrow] ($(phi1)+(.8cm,0)$) -- ($(phi1)+(1.4cm,0)$);

    \node[ps=2,inner ysep=.15cm,rectangle split empty part width=.16cm,fill=mygreen] (queue2) at (queue2) {};
    \fill[white] ([xshift=-\pgflinewidth-1pt,yshift=-\pgflinewidth+1pt] queue2.north east)
    rectangle ([xshift=\pgflinewidth+4pt,yshift=\pgflinewidth-1pt] queue2.north west);
    \fill[white] ([xshift=-\pgflinewidth-1pt,yshift=-\pgflinewidth] queue2.north east)
    rectangle ([xshift=15pt,yshift=\pgflinewidth] queue2.north west);
    \draw[-] ([xshift=15pt] queue2.north east) -- ([xshift=15pt] queue2.north west);

    \node[serverps] (phi2) at ($(queue2.south)+(.55cm,0)$) {};
    \node at (phi2) {$\phi_2(x)$};
    \node[anchor=south] at ($(queue2)+(.08cm,.35cm)$) {$x_2 = 2$};

    \draw[arrow] ($(queue2.north)+(-.8cm,0)$) -- node[pos=-.1,anchor=east] {$\lambda_2$} ($(queue2.north)-(.2cm,0)$);
    \draw[arrow] ($(phi2)+(.8cm,0)$) -- ($(phi2)+(1.4cm,0)$);
  \end{tikzpicture}
  \caption{An open Whittle network of $N = 2$ queues associated with the server pool of \figurename~\ref{fig:bipartite}.}
  \label{fig:whittle}
\end{figure}

Since all jobs of the same class receive service at the same rate,
we can describe the evolution of the system
with a network of $N$ \gls{ps} queues with state-dependent service capacities.
For each $i \in \classes$,
queue $i$ contains jobs of class $i$;
the arrival rate at this queue is $\lambda_i$
and its service capacity is $\phi_i(x)$ when the network state is $x$.
An example is shown in \figurename~\ref{fig:whittle} for the configuration of \figurename~\ref{fig:bipartite}.

\paragraph*{Capacity set}

The compatibilities between classes and servers
restrict the set of feasible resource allocations.
Specifically,
the vector $(\phi_i(x) : i \in \classes)$ of per-class service rates
belongs to the following capacity set in any state $x \in \states$:
$$
\arrivals = \left\{
  \phi \in \R_+^N:~
  \sum_{i \in \A} \phi_i \le \mu(\A),~
  \forall \A \subset \classes
\right\}.
$$
As observed in \cite{SV15},
the properties
satisfied by $\mu$ guarantee that
$\arrivals$ is a polymatroid \cite{F05}.

\paragraph*{Balance function}

It was shown in \cite{BP02} that
the resource allocation is insensitive if and only if
there is a balance function $\Phi$ defined on $\states$
such that $\Phi(0) = 1$ and
\begin{equation}
  \label{eq:phiPhi}
  \phi_i(x) = \frac{\Phi(x-e_i)}{\Phi(x)},
  \quad \forall x \in \states,
  \quad \forall i \in \classes(x),
\end{equation}
where $e_i$ is the $N$-dimensional vector with $1$ in component $i$ and $0$ elsewhere
and $\classes(x) = \{i \in \classes: x_i > 0\}$ is the set of active classes in state $x$.
Under this condition,
the network of \gls{ps} queues defined above is a Whittle network \cite{S99}.
The insensitive resource allocations that respect the capacity constraints of the system
are characterized by a balance function $\Phi$ such that, for all $x \in \states \setminus \{0\}$,
\begin{equation*}
  \Phi(x) \ge \frac1{\mu(\A)} \sum_{i \in \A} \Phi(x-e_i),
  \quad \forall \A \subset \classes(x),~\A \neq \emptyset.
\end{equation*}
Recursively maximizing the overall service rate in the system
is then equivalent to minimizing $\Phi$ by choosing
\begin{equation*}
  \Phi(x) = \max_{\substack{\A \subset \classes(x), \\ \A \neq \emptyset}} \left(
    \frac1{\mu(\A)} \sum_{i \in \A} \Phi(x-e_i)
  \right),
  \quad \forall x \in \states \setminus \{0\}.
\end{equation*}
The resource allocation defined by this balance function is called balanced fairness.

It was shown in \cite{SV15} that balanced fairness is Pareto-efficient in polymatroid capacity sets,
meaning that the total service rate $\sum_{i \in \classes(x)} \phi_i(x)$ is always equal to
the aggregate capacity $\mu(\classes(x))$ of the servers that can process at least one active class.
By \eqref{eq:phiPhi}, this is equivalent to
\begin{equation}
  \label{eq:recPhix}
  \Phi(x) = \frac1{\mu(\classes(x))} \sum_{i \in \classes(x)} \Phi(x-e_i),
  \quad \forall x \in \states \setminus \{0\}.
\end{equation}

\paragraph*{Stationary distribution}

The Markov process defined by the system state $x$ is reversible, with stationary distribution
\begin{equation}
  \label{eq:defpix}
  \pi(x) = \pi(0) \Phi(x) \prod_{i \in \classes} {\lambda_i}^{x_i},
  \quad \forall x \in \states.
\end{equation}
By insensitivity,
the system state has the same stationary distribution
if the jobs sizes within each class are only i.i.d.,
as long as the traffic intensity of class $i$
(defined as the average quantity of work brought by jobs of this class per unit of time)
is $\lambda_i$, for each $i \in \classes$.
A proof of this result is given in \cite{BP02} for Cox distributions,
which form a dense subset within the set of distributions of nonnegative random variables.

\subsection{Job scheduling}
\label{subsec:scheduling}

We now describe the sequential implementation of balanced fairness that was proposed in \cite{BC17}.
This will lay the foundations for the results of Section \ref{sec:load-balancing}.

We still assume that a job can be distributed among several servers,
but we relax the assumption that servers can process several jobs at the same time.
Instead, each server processes its jobs sequentially in \gls{fcfs} order.
When a job arrives,
it enters in service on every idle server within its assignment, if any,
so that its service rate is the sum of the capacities of these servers.
When the service of a job is complete,
it leaves the system immediately
and its servers are reallocated to the first job they can serve in the queue.
Note that this sequential implementation
also makes sense in a model where jobs are
\emph{replicated} over several servers instead of being processed \emph{in parallel}.
For more details, we refer the reader to \cite{G15}~where the model with redundant requests was introduced.

Since the arrival order of jobs impacts the rate allocation,
we need to detail the system state.
We consider the sequence $c = (c_1,\ldots,c_n) \in \classes^*$,
where $n$ is the number of jobs in the system
and $c_p$ is the class of the $p$-th oldest job, for each $p = 1,\ldots,n$.
$\emptyset$ denotes the empty state, with $n = 0$.
The vector of numbers of jobs of each class in the system,
corresponding to the state introduced in \S \ref{subsec:bf},
is denoted by $|c| = (|c|_i : i \in \classes) \in \states$.
It does not define a Markov process in general.
We let $\classes(c) = \classes(|c|)$
denote the set of active classes in state $c$.
The state space of this detailed system state is
$\seq = \{c \in \classes^*: |c| \le \ell\}$.

\paragraph*{Queueing model}

\begin{figure}[h]
  \centering

  \begin{tikzpicture}
    \node[fcfs=7,
      rectangle split part fill={white,white,mygreen,myorange,mygreen,myorange,myorange},
    ] (queue) {
      \nodepart {one} {}
      \nodepart {two} {}
      \nodepart {three} {2}
      \nodepart {four} {1}
      \nodepart {five} {2}
      \nodepart {six} {1}
      \nodepart {seven} {1}
    };
    \fill[white] ([xshift=-\pgflinewidth-1pt,yshift=-\pgflinewidth+1pt]queue.north west)
    rectangle ([xshift=\pgflinewidth+4pt,yshift=\pgflinewidth-1pt]queue.south west);
    \fill[white] ([xshift=-\pgflinewidth-1pt,yshift=-\pgflinewidth-.045pt]queue.north west)
    rectangle ([xshift=\pgflinewidth+10pt,yshift=\pgflinewidth+.045pt]queue.south west);
    \node[anchor=south,align=center] at ($(queue)+(.1cm,.35cm)$) {$c = (1,1,2,1,2)$};

    \node[serveroi] (mu) at ($(queue.east)+(.65cm,0)$) {};
    \node at (mu) {$\mu(\classes(c))$};

    \draw[arrow] ($(queue.west)+(-.8cm,+.22cm)$) -- node[pos=-.1,anchor=east] {$\lambda_1$} ($(queue.west)+(-.2cm,.22cm)$);
    \draw[arrow] ($(queue.west)+(-.8cm,-.22cm)$) -- node[pos=-.1,anchor=east] {$\lambda_2$} ($(queue.west)-(.2cm,.22cm)$);
    \draw[arrow] ($(mu)+(.9cm,0)$) -- ($(mu)+(1.5cm,0)$);

  \end{tikzpicture}
  \caption{
    An \gls{oi} queue with $N = 2$ job classes associated with the server pool of \figurename~\ref{fig:bipartite}.
    The job of class $1$ at the head of the queue is in service on servers $1$ and $2$.
    The third job, of class $2$, is in service on server $3$.
    Aggregating the state $c$ yields the state $x$ of the Whittle network of \figurename~\ref{fig:whittle}.
  }
  \label{fig:oi-open}
\end{figure}
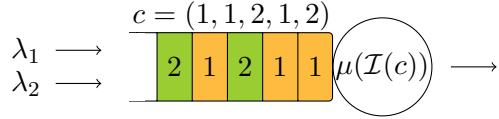

Each job is in service on all the servers
that were assigned this job
but not those that arrived earlier.
For each $p = 1,\ldots,n$,
the service rate of the job in position $p$
is thus given by
\begin{equation*}
  \sum_{s \in \servers_{c_p} \setminus \bigcup_{q=1}^{p-1} \servers_{c_q}} \mu_s
  = \mu(\classes(c_1,\ldots,c_p)) - \mu(\classes(c_1,\ldots,c_{p-1})),
\end{equation*}
with the convention that $(c_1,\ldots,c_{p-1}) = \emptyset$ if $p = 1$.
The service rate of a job
is independent of the jobs arrived later in the system.
Additionally,
the total service rate $\mu(\classes(c))$
is independent of the arrival order of jobs.
The corresponding queueing model is an \gls{oi} queue \cite{BK96,K11}.
An example is shown in \figurename~\ref{fig:oi-open} for the configuration of \figurename~\ref{fig:bipartite}.

\paragraph*{Stationary distribution}

The Markov process defined by the system state $c$ is irreducible.
The results of \cite{K11} show that
this process
is quasi-reversible, with stationary distribution
\begin{equation}
  \label{eq:defpic}
  \pi(c) = \pi(\emptyset) \Phi(c) \prod_{i \in \classes} {\lambda_i}^{|c|_i},
  \quad \forall c \in \seq,
\end{equation}
where $\Phi$ is defined recursively on $\seq$ by $\Phi(\emptyset) = 1$ and
\begin{equation}
  \label{eq:recPhic}
  \Phi(c) = \frac1{\mu(\classes(c))} \Phi(c_1,\ldots,c_{n-1}),
  \quad \forall c \in \seq \setminus \{\emptyset\}.
\end{equation}

We now go back to the aggregate state $x$ giving the number of jobs of each class in the system.
With a slight abuse of notation,
we let
\begin{equation*}
  \pi(x)
  = \sum_{c:|c|=x} \pi(c)
  \quad \text{and} \quad
  \Phi(x)
  = \sum_{c:|c|=x} \Phi(c),
  \quad \forall x \in \states.
\end{equation*}
As observed in \cite{K11,BC17},
if follows from \eqref{eq:defpic} that
\begin{align*}
  \pi(x)
  = \pi(\emptyset) \left( \sum_{c:|c|=x} \Phi(c) \right) \prod_{i \in \classes} {\lambda_i}^{x_i}
  = \pi(0) \Phi(x) \prod_{i \in \classes} {\lambda_i}^{x_i}
\end{align*}
in any state $x$.
Using \eqref{eq:recPhic}, we can show
that $\Phi$ satisfies \eqref{eq:recPhix}
with the initial condition $\Phi(0) = \Phi(\emptyset) = 1$.
Hence, the stationary distribution of the aggregate system state $x$
is exactly that obtained in \S \ref{subsec:bf} under balanced fairness.

It was also shown in \cite{BC17} that
the average per-class resource allocation resulting from \gls{fcfs} service discipline is balanced fairness.
In other words, we have
\begin{equation*}
  \phi_i(x) = \sum_{c:|c|=x} \frac{\pi(c)}{\pi(x)} \mu_i(c),
  \quad \forall x \in \states,
  \quad \forall i \in \classes(x),
\end{equation*}
where
$\phi_i(x)$ is the total service rate allocated to class-$i$ jobs in state $x$ under balanced fairness,
given by \eqref{eq:phiPhi},
and $\mu_i(c)$ denotes the service rate received by the first job of class $i$ in state $c$ under \gls{fcfs} service discipline:
\begin{equation*}
  \mu_i(c) = \sum_{\substack{p=1 \\ c_p = i}}^n 
  ( \mu(\classes(c_1,\ldots,c_p)) - \mu(\classes(c_1,\ldots,c_{p-1})) ).
\end{equation*}
Observe that, by \eqref{eq:defpix} and \eqref{eq:defpic}, the rate equality simplifies to
\begin{equation}
  \label{eq:phimu}
  \phi_i(x) = \sum_{c:|c|=x} \frac{\Phi(c)}{\Phi(x)} \mu_i(c),
  \quad \forall x \in \states,
  \quad \forall i \in \classes(x).
\end{equation}
We will use this last equality later.

As it is, the \gls{fcfs} service discipline is very sensitive to the job size distribution.
\cite{BC17}
mitigates this sensitivity by frequently interrupting jobs and moving them to the end of the queue,
in the same way as round-robin scheduling algorithm in the single-server case.
In the queueing model,
these interruptions and resumptions are represented approximately by random routing,
which leaves the stationary distribution unchanged
by quasi-reversibility \cite{kelly,S99}.
If the interruptions are frequent enough,
then all jobs of a class tend to receive
the same service rate on average,
which is that obtained under balanced fairness.
In particular, performance becomes approximately insensitive to the job size distribution within each class.

\section{Load balancing}
\label{sec:load-balancing}

The previous section has considered the problem of resource sharing.
We now focus on dynamic load balancing,
using the fact that each job may be \textit{a priori} compatible with several classes
and assigned to one of them upon arrival.
We first extend the model of \S \ref{subsec:bipartite} to add this new degree of flexibility.

\subsection{Model}
\label{subsec:tripartite}

We again consider a pool of $S$ servers.
There are $N$ job classes
and we let $\classes = \{1,\ldots,N\}$ denote the set of class indices.
The compatibilities between job classes and servers are described by a bipartite graph, as explained in \S \ref{subsec:bipartite}.
Additionally, we assume that the arrivals are divided into $K$ types,
so that the jobs of each type enter the system according to an independent Poisson process.
Job sizes are independent and exponentially distributed with unit mean.
Each job leaves the system immediately after service completion.

The type of a job defines the set of classes it can be assigned to.
This assignment is performed instantaneously upon the job arrival,
according to some decision rule that will be detailed later.
For each $i \in \classes$,
we let $\types_i \subset \{1,\ldots,K\}$
denote the non-empty set of job types that can be assigned to class $i$.
This defines a bipartite compatibility graph between types and classes,
where there is an edge between a type and a class if the jobs of this type can be assigned to this class.
Overall, the compatibilities are described by a tripartite graph
between types, classes, and servers.
\figurename~\ref{fig:tripartite} shows a toy example.

For each $k = 1,\ldots,K$,
the arrival rate of type-$k$ jobs in the system is denoted by $\nu_k > 0$.
We can then define a function $\nu$ on the power set of $\classes$ as follows:
for each $\A \subset \classes$,
$$
\nu(\A) = \sum_{k \in \bigcup_{i \in \A} \types_i} \nu_k
$$
denotes the aggregate arrival rate of the types
that can be assigned to at least one class in $\A$.
$\nu$ satisfies the submodularity, monotonicity and normalization properties
satisfied by the function $\mu$ of \S \ref{subsec:bipartite}.

\subsection{Randomized load balancing}
\label{subsec:random}

We now express the insensitive load balancing of \cite{BJP04}
in our new server pool model.
This extends the static load balancing considered earlier.
Incoming jobs are assigned to classes at random,
and the assignment probabilities depend
not only on the job type but also on the system state.
As in \S \ref{subsec:bf},
we assume that the capacity of each server
can be divided continuously between its jobs.
The resources are allocated by applying balanced fairness
in the capacity set defined by the bipartite compatibility graph between job classes and servers.

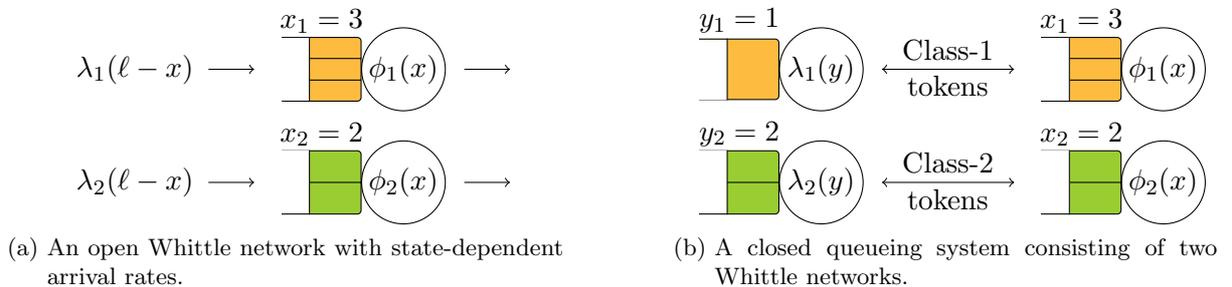
\begin{figure}[b]
  \centering
  \subfloat[An open Whittle network with state-dependent arrival rates.\label{fig:whittle-open}]{
    \begin{tikzpicture}
      \node (queue1) {};
      \node (queue2) at ($(queue1)+(0,-1.5cm)$) {};

      \node[ps=3,inner ysep=.15cm,rectangle split empty part width=.02cm,fill=myorange] (queue1) at (queue1) {};
      \fill[white] ([xshift=-\pgflinewidth-1pt,yshift=-\pgflinewidth+1pt] queue1.north east)
      rectangle ([xshift=\pgflinewidth+4pt,yshift=\pgflinewidth-1pt] queue1.north west);
      \fill[white] ([xshift=-\pgflinewidth-1pt,yshift=-\pgflinewidth] queue1.north east)
      rectangle ([xshift=15pt,yshift=\pgflinewidth] queue1.north west);
      \draw[-] ([xshift=15pt] queue1.north east) -- ([xshift=15pt] queue1.north west);

      \node[serverps] (phi1) at ($(queue1.south)+(.55cm,0)$) {};
      \node at (phi1) {$\phi_1(x)$};
      \node[anchor=south] at ($(queue1)+(.08cm,.35cm)$) {$x_1 = 3$};

      \node[ps=2,inner ysep=.15cm,rectangle split empty part width=.16cm,fill=mygreen] (queue2) at (queue2) {};
      \fill[white] ([xshift=-\pgflinewidth-1pt,yshift=-\pgflinewidth+1pt] queue2.north east)
      rectangle ([xshift=\pgflinewidth+4pt,yshift=\pgflinewidth-1pt] queue2.north west);
      \fill[white] ([xshift=-\pgflinewidth-1pt,yshift=-\pgflinewidth] queue2.north east)
      rectangle ([xshift=15pt,yshift=\pgflinewidth] queue2.north west);
      \draw[-] ([xshift=15pt] queue2.north east) -- ([xshift=15pt] queue2.north west);

      \node[serverps] (phi2) at ($(queue2.south)+(.55cm,0)$) {};
      \node at (phi2) {$\phi_2(x)$};
      \node[anchor=south] at ($(queue2)+(.08cm,.35cm)$) {$x_2 = 2$};

      \draw[arrow] ($(queue1.north)+(-.8cm,0)$) -- node[pos=-.1,anchor=east] {$\lambda_1(\ell-x)$} ($(queue1.north)-(.2cm,0)$);
      \draw[arrow] ($(phi1)+(.8cm,0)$) -- ($(phi1)+(1.4cm,0)$);
      \draw[arrow] ($(queue2.north)+(-.8cm,0)$) -- node[pos=-.1,anchor=east] {$\lambda_2(\ell-x)$} ($(queue2.north)-(.2cm,0)$);
      \draw[arrow] ($(phi2)+(.8cm,0)$) -- ($(phi2)+(1.4cm,0)$);

      \node at (-3.8cm,0) {};
      \node at (3cm,0) {};
    \end{tikzpicture}
  }
  \hfill
  \subfloat[A closed queueing system consisting of two Whittle networks.
  \label{fig:whittle-closed}]{
    \begin{tikzpicture}

      \node (queue1)  {};
      \node (queue2) at ($(queue1)+(0,-1.5cm)$) {};
      \node (arrival1) at ($(queue1)-(4.5cm,0)$) {};
      \node (arrival2) at ($(arrival1)+(0,-1.5cm)$) {};

      \node[ps=1,inner ysep=.15cm, rectangle split empty part width=.55cm,fill=myorange] (arrival1) at (arrival1) {};
      \fill[white] ([xshift=-\pgflinewidth-1pt,yshift=-\pgflinewidth+1pt] arrival1.north east)
      rectangle ([xshift=\pgflinewidth+4pt,yshift=\pgflinewidth-1pt] arrival1.north west);
      \fill[white] ([xshift=-\pgflinewidth-1pt,yshift=-\pgflinewidth] arrival1.north east)
      rectangle ([xshift=15pt,yshift=\pgflinewidth] arrival1.north west);
      \draw[-] ([xshift=15pt] arrival1.north east) -- ([xshift=15pt] arrival1.north west);

      \node[serverps] (lambda1) at ($(arrival1.south)+(.55cm,0)$) {};
      \node at (lambda1) {$\lambda_1(y)$};
      \node[anchor=south] at ($(arrival1)+(.08cm,.35cm)$) {$y_1 = 1$};

      \node[ps=2,inner ysep=.15cm,rectangle split empty part width=.16cm,fill=mygreen] (arrival2) at (arrival2) {};
      \fill[white] ([xshift=-\pgflinewidth-1pt,yshift=-\pgflinewidth+1pt] arrival2.north east)
      rectangle ([xshift=\pgflinewidth+4pt,yshift=\pgflinewidth-1pt] arrival2.north west);
      \fill[white] ([xshift=-\pgflinewidth-1pt,yshift=-\pgflinewidth] arrival2.north east)
      rectangle ([xshift=15pt,yshift=\pgflinewidth] arrival2.north west);
      \draw[-] ([xshift=15pt] arrival2.north east) -- ([xshift=15pt] arrival2.north west);

      \node[serverps] (lambda2) at ($(arrival2.south)+(.55cm,0)$) {};
      \node at (lambda2) {$\lambda_2(y)$};
      \node[anchor=south] at ($(arrival2)+(.08cm,.35cm)$) {$y_2 = 2$};

      \node[ps=3,inner ysep=.15cm,rectangle split empty part width=.02cm,fill=myorange] (queue1) at (queue1) {};
      \fill[white] ([xshift=-\pgflinewidth-1pt,yshift=-\pgflinewidth+1pt] queue1.north east)
      rectangle ([xshift=\pgflinewidth+4pt,yshift=\pgflinewidth-1pt] queue1.north west);
      \fill[white] ([xshift=-\pgflinewidth-1pt,yshift=-\pgflinewidth] queue1.north east)
      rectangle ([xshift=15pt,yshift=\pgflinewidth] queue1.north west);
      \draw[-] ([xshift=15pt] queue1.north east) -- ([xshift=15pt] queue1.north west);

      \node[serverps] (phi1) at ($(queue1.south)+(.55cm,0)$) {};
      \node at (phi1) {$\phi_1(x)$};
      \node[anchor=south] at ($(queue1)+(.08cm,.35cm)$) {$x_1 = 3$};

      \node[ps=2,inner ysep=.15cm,rectangle split empty part width=.16cm,fill=mygreen] (queue2) at (queue2) {};
      \fill[white] ([xshift=-\pgflinewidth-1pt,yshift=-\pgflinewidth+1pt] queue2.north east)
      rectangle ([xshift=\pgflinewidth+4pt,yshift=\pgflinewidth-1pt] queue2.north west);
      \fill[white] ([xshift=-\pgflinewidth-1pt,yshift=-\pgflinewidth] queue2.north east)
      rectangle ([xshift=15pt,yshift=\pgflinewidth] queue2.north west);
      \draw[-] ([xshift=15pt] queue2.north east) -- ([xshift=15pt] queue2.north west);

      \node[serverps] (phi2) at ($(queue2.south)+(.55cm,0)$) {};
      \node at (phi2) {$\phi_2(x)$};
      \node[anchor=south] at ($(queue2)+(.08cm,.35cm)$) {$x_2 = 2$};

      \draw[arrow,<->] ($(lambda1)+(.8cm,0)$) --
      node[midway,align=center,yshift=.02cm] {Class-$1$ \\ tokens}
      ($(queue1.north)-(.2cm,0)$);
      \draw[arrow,<->] ($(lambda2)+(.8cm,0)$) --
      node[midway,align=center,yshift=.02cm] {Class-$2$ \\ tokens}
      ($(queue2.north)-(.2cm,0)$);

    \end{tikzpicture}
  }
  \caption{
    Alternative representations of a Whittle network associated with the server pool of \figurename~\ref{fig:tripartite}.
    At most $\ell_1 = \ell_2 = 4$ jobs can be assigned to each class.
  }
  \label{fig:whittle-dlb}
\end{figure}

\paragraph*{Open queueing model}

We first recall the queueing model considered in \cite{BJP04} to describe the randomized load balancing.
As in \S \ref{subsec:bf},
jobs are gathered by class in \gls{ps} queues with state-dependent service capacities given by \eqref{eq:phiPhi}.
Hence, the type of a job is forgotten once it is assigned to a class.

Similarly, we record the job arrivals depending on the class they are assigned to,
regardless of their type before the assignment.
The Poisson arrival assumption ensures that,
given the system state,
the time before the next arrival at each class
is exponentially distributed and independent of the arrivals at other classes.
The rates of these arrivals result from the load balancing.
We write them as functions of the vector $y = \ell - x$
of numbers of available positions at each class.
Specifically,
$\lambda_i(y)$
denotes the arrival rate of jobs assigned to class $i$ when
there are $y_j$ available positions in class $j$,
for each $j \in \classes$.

The system can thus be modeled by a network of $N$ \gls{ps} queues with state-dependent arrival rates,
as shown in \figurename~\ref{fig:whittle-open}.

\paragraph*{Closed queueing model}

We introduce a second queueing model
that describes the system dynamics differently.
It will later simplify the study of the insensitive load balancing
by drawing a parallel with the resource allocation of \S \ref{subsec:bf}.

Our alternative model stems from the following observation:
since we impose limits on the number of jobs of each class,
we can indifferently assume that the arrivals are
limited by the intermediary of buckets containing tokens.
Specifically, for each $i \in \classes$,
the assignments to class $i$ are controlled
through a bucket filled with $\ell_i$ tokens.
A job that is assigned to class $i$ removes a token from this bucket
and holds it until its service is complete.
The assignments to a class are suspended when the bucket of this class is empty,
and they are resumed when a token of this class is released.

Each token is either
held by a job in service or waiting to be seized by an incoming job.
We consider a closed queueing model that reflects this alternation:
a first network of $N$ queues contains tokens held by jobs in service, as before,
and a second network of $N$ queues contains available tokens.
For each $i \in \classes$,
a token of class $i$ alternates between the queues indexed by $i$ in the two networks.
This is illustrated in \figurename~\ref{fig:whittle-closed}.

The state of the network containing tokens held by jobs in service is $x$.
The queues in this network apply \gls{ps} service discipline
and their service capacities are given by \eqref{eq:phiPhi}.
The state of the network containing available tokens is $y = \ell - x$.
For each $i \in \classes$,
the service of a token at queue $i$ in this network
is triggered by the arrival of a job assigned to class $i$.
The service capacity of this queue is thus equal to $\lambda_i(y)$ in state $y$.
Since all tokens of the same class are exchangeable,
we can assume indifferently that we pick one of them at random,
so that the service discipline of the queue is \gls{ps}.

\paragraph*{Capacity set}

The compatibilities between job types and classes
restrict the set of feasible load balancings.
Specifically,
the vector $(\lambda_i(y): i \in \classes)$ of per-class arrival rates
belongs to the following capacity set in any state $y \in \states$:
\begin{equation*}
  \departures = \left\{
    \lambda \in \R_+^N:~
    \sum_{i \in \A} \lambda_i \le \nu(\A),~
    \forall \A \subset \classes
  \right\}.
\end{equation*}
The properties satisfied by $\nu$
guarantee that $\departures$ is a polymatroid.

\paragraph*{Balance function}

Our token-based reformulation allows us to interpret dynamic load balancing
as a problem of resource allocation in the network of queues containing available tokens.
This will allow us to apply the results of \S \ref{subsec:bf}.

It was shown in \cite{BJP04} that
the load balancing is insensitive
if and only if
there is a balance function $\Lambda$ defined on $\states$ such that
$\Lambda(0) = 1$, and
\begin{align}
  \label{eq:lambdaLambda}
  \lambda_i(y) = \frac{\Lambda(y-e_i)}{\Lambda(y)},
  \quad \forall y \in \states,
  \quad \forall i \in \classes(y).
\end{align}
Under this condition, the network of \gls{ps} queues containing available tokens
is a Whittle network.

The Pareto-efficiency of balanced fairness in polymatroid capacity sets
can be understood as follows in terms of load balancing.
We consider the balance function $\Lambda$ defined recursively on $\states$ by $\Lambda(0) = 1$ and
\begin{equation}
  \label{eq:recLambday}
  \Lambda(y) = \frac1{\nu(\classes(y))} \sum_{i \in \classes(y)} \Lambda(y-e_i),
  \quad \forall y \in \states \setminus \{0\}.
\end{equation}
Then $\Lambda$ defines a load balancing
that belongs to the capacity set $\departures$ in each state $y$.
By \eqref{eq:lambdaLambda}, this load balancing satisfies
\begin{equation*}
  \sum_{i \in \classes(y)} \lambda_i(y) = \nu(\classes(y)),
  \quad \forall y \in \states,
\end{equation*}
meaning that an incoming job is accepted
whenever it is compatible with at least one available token.

\paragraph*{Stationary distribution}

The Markov process defined by the system state $x$ is reversible,
with stationary distribution
\begin{equation}
  \label{eq:stationary}
  \pi(x) = \frac1G \Phi(x) \Lambda(\ell-x),
  \quad \forall x \in \states,
\end{equation}
where $G$ is a normalization constant.
Note that we could symmetrically give the stationary distribution
of the Markov process defined by the vector $y = \ell-x$ of numbers of available tokens.
As mentioned earlier,
the insensitivity of balanced fairness is preserved by the load balancing.

\subsection{Deterministic token mechanism}
\label{subsec:deterministic}

Our closed queueing model reveals
that the randomized load balancing
is dual to the balanced fair resource allocation.
This allows us to propose a new \emph{deterministic} load balancing algorithm
that mirrors the \gls{fcfs} service discipline of \S \ref{subsec:scheduling}.
This algorithm can be combined indifferently
with balanced fairness or with the sequential \gls{fcfs} scheduling;
in both cases,
we show that it implements the load balancing defined by \eqref{eq:lambdaLambda}.

All available tokens are now sorted
in order of release in a single bucket.
The longest available tokens are in front.
An incoming job scans the bucket from beginning to end
and seizes the first compatible token;
it is blocked if it does not find any.
For now, we assume that the server resources are allocated to the accepted jobs
by applying the \gls{fcfs} service discipline of \S \ref{subsec:scheduling}.
When the service of a job is complete,
its token is released and added to the end of the bucket.

We describe the system state with a couple $(c,t)$
retaining both the arrival order of jobs
and the release order of tokens.
Specifically,
$c = (c_1,\ldots,c_n) \in \seq$
is the sequence of classes of (tokens held by) jobs in service, as before,
and $t = (t_1,\ldots,t_m) \in \seq$
is the sequence of classes of available tokens, ordered by release,
so that $t_1$ is the class of the longest available token.
Given the total number of tokens of each class in the system,
any feasible state satisfies $|c| + |t| = \ell$.

\paragraph*{Queueing model}

Depending on its position in the bucket,
each available token is seized by any incoming job
whose type is compatible with this token
but not with the tokens released earlier.
For each $p = 1,\ldots,m$,
the token in position $p$ is thus seized at rate
\begin{equation*}
  \sum_{k \in \types_{t_p} \setminus \bigcup_{q=1}^{p-1} \types_{t_q}} \nu_k
  = \nu(\classes(t_1,\ldots,t_p)) - \nu(\classes(t_1,\ldots,t_{p-1})).
\end{equation*}

The seizing rate of a token
is independent of the tokens released later.
Additionally, the total rate at which available tokens are seized is $\nu(\classes(y))$,
independently of their release order.
The bucket can thus be modeled by an \gls{oi} queue,
where the service of a token is triggered by the arrival of a job that seizes this token.

\begin{figure}[h]
  \centering

  \begin{tikzpicture}
    \node[fcfs=7,
      rectangle split part fill={white,white,mygreen,myorange,mygreen,myorange,myorange},
    ] (queue) {
      \nodepart {one} {}
      \nodepart {two} {}
      \nodepart {three} {2}
      \nodepart {four} {1}
      \nodepart {five} {2}
      \nodepart {six} {1}
      \nodepart {seven} {1}
    };
    \fill[white] ([xshift=-\pgflinewidth-1pt,yshift=-\pgflinewidth+1pt]queue.north west)
    rectangle ([xshift=\pgflinewidth+4pt,yshift=\pgflinewidth-1pt]queue.south west);
    \fill[white] ([xshift=-\pgflinewidth-1pt,yshift=-\pgflinewidth-.045pt]queue.north west)
    rectangle ([xshift=\pgflinewidth+10pt,yshift=\pgflinewidth+.045pt]queue.south west);
    \node[anchor=south] at ($(queue)+(.1cm,.35cm)$) {$c = (1,1,2,1,2)$};

    \node[serveroi] (mu) at ($(queue.east)+(.65cm,0)$) {};
    \node at (mu) {$\mu(\classes(c))$};

    \node[fcfs=7,
      rectangle split part fill={myorange,mygreen,mygreen,white},
    ] (bucket) at ($(queue)+(1.3cm,1.8cm)$) {
      \nodepart {one} {1}
      \nodepart {two} {2}
      \nodepart {three} {2}
      \nodepart {four} {\enspace}
      \nodepart {five} {\enspace}
      \nodepart {six} {}
      \nodepart {seven} {}
    };
    \fill[white] ([xshift=\pgflinewidth+1pt,yshift=-\pgflinewidth+1pt]bucket.north east)
    rectangle ([xshift=-\pgflinewidth-4pt,yshift=\pgflinewidth-1pt]bucket.south east);
    \fill[white] ([xshift=\pgflinewidth+1pt,yshift=-\pgflinewidth-.045pt]bucket.north east)
    rectangle ([xshift=-\pgflinewidth-10pt,yshift=\pgflinewidth+.045pt]bucket.south east);
    \node[anchor=south] at ($(bucket)+(.1cm,.35cm)$) {$t = (1,2,2)$};

    \node[serveroi] (nu) at ($(bucket.west)-(.65cm,0)$) {};
    \node at (nu) {$\nu(\classes(t))$};

    \draw[arrow] ($(mu)+(.9cm,0)$) -- ($(mu)+(1.4cm,0)$) |- ($(bucket.east)+(.2cm,0)$);
    \draw[arrow] ($(nu)-(.9cm,0)$) -| ($(queue.west)+(-.8cm,0)$) -- ($(queue.west)+(-.2cm,0)$);
  \end{tikzpicture}
  \caption{
    A closed tandem network of two \gls{oi} queues associated with the server pool of \figurename~\ref{fig:tripartite}.
    At most $\ell_1 = \ell_2 = 4$ jobs can be assigned to each class.
    The state is $(c,t)$,
    with $c = (1,1,2,1,2)$ and $t = (1,2,2)$.
    The corresponding aggregate state is that of the network of \figurename~\ref{fig:whittle-dlb}.
    An incoming job of type $1$ would seize the available token in first position (of class $1$),
    while an incoming job of type $2$ would seize the available token in second position (of class $2$).
  }
  \label{fig:oi-tandem}
\end{figure}
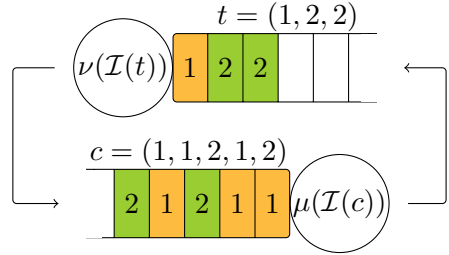

The evolution of the sequence of tokens held by jobs in service also defines an \gls{oi} queue,
with the same dynamics as in \S \ref{subsec:scheduling}.
Overall, the system can be modeled by a closed tandem network of two \gls{oi} queues,
as shown in \figurename~\ref{fig:oi-tandem}.

\paragraph*{Stationary distribution}

Assuming $\servers_i \neq \servers_j$ or $\types_i \neq \types_j$
for each pair $\{i,j\} \subset \classes$ of classes,
the Markov process defined by
the detailed state $(c,t)$ is irreducible.
The proof is provided in the appendix.
Known results on networks of quasi-reversible queues \cite{kelly}
then show that this process
is quasi-reversible, with stationary distribution
\begin{equation*}
  \pi(c,t) = \frac1G \Phi(c) \Lambda(t),
  \quad \forall c,t \in \seq: |c| + |t| = \ell,
\end{equation*}
where $\Phi$ is defined by the recursion \eqref{eq:recPhic} and the initial step $\Phi(\emptyset) = 1$, as in \S \ref{subsec:scheduling};
similarly, $\Lambda$ is defined recursively on $\seq$ by $\Lambda(\emptyset) = 1$ and
\begin{equation*}
  \Lambda(t) = \frac1{\nu(\classes(t))} \Lambda(t_1,\ldots,t_{m-1}),
  \quad \forall t \in \seq \setminus \{\emptyset\}.
\end{equation*}

We go back to the aggregate state $x$ giving the number of tokens of each class held by jobs in service.
With a slight abuse of notation,
we define its stationary distribution by
\begin{equation}
  \label{eq:defpict}
  \pi(x) = \sum_{c:|c|=x}\,\sum_{t:|t|=\ell-x} \pi(c,t),
  \quad \forall x \in \states.
\end{equation}
As in \S \ref{subsec:scheduling},
we can show that we have
\begin{equation*}
  \pi(x) = \frac1G \Phi(x) \Lambda(\ell-x),
  \quad \forall x \in \states,
\end{equation*}
where the functions $\Phi$ and $\Lambda$ are defined on $\states$ by
\begin{align*}
  \Phi(x) = \sum_{c:|c|=x} \Phi(c)
  \quad \text{and} \quad
  \Lambda(y) = \sum_{t:|t|=y} \Lambda(t),
  \quad \forall x,y \in \states,
\end{align*}
respectively.
These functions $\Phi$ and $\Lambda$
satisfy the recursions \eqref{eq:recPhix} and \eqref{eq:recLambday}, respectively,
with the initial conditions $\Phi(0) = \Lambda(0) = 1$.
Hence, the aggregate stationary distribution of the system state $x$
is exactly that obtained in \S \ref{subsec:random}
by combining the randomized load balancing with balanced fairness.

Also, using the definition of $\Lambda$,
we can rewrite \eqref{eq:phimu} as follows:
for each $x \in \states$ and $i \in \classes(x)$,
\begin{align*}
  \phi_i(x)
  &= \sum_{c:|c|=x} \frac{\frac1G \Phi(c) \sum_{t:|t|=\ell-x} \Lambda(t)}{\frac1G \Phi(x) \Lambda(\ell-x)} \mu_i(c), \\
  &= \sum_{c:|c|=x}\,\sum_{t:|t|=\ell-x} \frac{\pi(c,t)}{\pi(x)} \mu_i(c).
\end{align*}
Hence, the average per-class service rates are still as defined by balanced fairness.
By symmetry, it follows that the average per-class arrival rates,
ignoring the release order of tokens,
are as defined by the randomized load balancing.
Specifically,
for each $y \in \states$ and $i \in \classes(y)$,
we have
\begin{equation*}
  \lambda_i(y) = \sum_{c:|c|=\ell-y}\,\sum_{t:|t|=y} \frac{\pi(c,t)}{\pi(\ell-y)} \nu_i(t),
\end{equation*}
where
$\lambda_i(y)$ is the arrival rate of jobs assigned to class $i$ in state $y$ under the randomized load balancing,
given by \eqref{eq:lambdaLambda},
and $\nu_i(t)$ denotes the rate at which the first available token of class $i$ is seized
under the deterministic load balancing:
\begin{equation*}
  \nu_i(t) = \sum_{\substack{p=1 \\ t_p = i}}^m
  ( \nu(\classes(t_1,\ldots,t_p)) - \nu(\classes(t_1,\ldots,t_{p-1})) ).
\end{equation*}

As in \S \ref{subsec:scheduling},
the stationary distribution of the system state is unchanged
by the addition of random routing,
as long as the average traffic intensity of each class remains constant.
Hence we can again reach some approximate insensitivity to the job size distribution within each class
by enforcing frequent job interruptions and resumptions.

\paragraph*{Application with balanced fairness}

\begin{figure}[h]
  \centering
  \begin{tikzpicture}

    \node[fcfs=5,
      rectangle split part fill={white,white,mygreen,mygreen,myorange},
    ] (arrival) {
      \nodepart {one} {}
      \nodepart {two} {}
      \nodepart {three} {2}
      \nodepart {four} {2}
      \nodepart {five} {1}
    };
    \fill[white] ([xshift=-\pgflinewidth-1pt,yshift=-\pgflinewidth+1pt]arrival.north west)
    rectangle ([xshift=\pgflinewidth+4pt,yshift=\pgflinewidth-1pt]arrival.south west);
    \fill[white] ([xshift=-\pgflinewidth-1pt,yshift=-\pgflinewidth-.06pt]arrival.north west)
    rectangle ([xshift=\pgflinewidth+10pt,yshift=\pgflinewidth+.06pt]arrival.south west);
    \node[anchor=south,align=center] at ($(arrival)+(.1cm,.35cm)$)
    {$t = (1,2,2)$};

    \node[serveroi] (nu) at ($(arrival.east)+(.65cm,0)$) {};
    \node at (nu) {$\nu(\classes(t))$};

    \node (queue1) at ($(nu)+(3.5cm,.75cm)$) {};
    \node (queue2) at ($(queue1)+(0,-1.5cm)$) {};

    \node[ps=3,inner ysep=.15cm,rectangle split empty part width=.02cm,fill=myorange] (queue1) at (queue1) {};
    \fill[white] ([xshift=-\pgflinewidth-1pt,yshift=-\pgflinewidth+1pt] queue1.north east)
    rectangle ([xshift=\pgflinewidth+4pt,yshift=\pgflinewidth-1pt] queue1.north west);
    \fill[white] ([xshift=-\pgflinewidth-1pt,yshift=-\pgflinewidth] queue1.north east)
    rectangle ([xshift=15pt,yshift=\pgflinewidth] queue1.north west);
    \draw[-] ([xshift=15pt] queue1.north east) -- ([xshift=15pt] queue1.north west);

    \node[serverps] (phi1) at ($(queue1.south)+(.55cm,0)$) {};
    \node at (phi1) {$\phi_1(x)$};
    \node[anchor=south] at ($(queue1)+(0,.35cm)$) {$x_1 = 3$};

    \node[ps=2,inner ysep=.15cm,rectangle split empty part width=.16cm,fill=mygreen] (queue2) at (queue2) {};
    \fill[white] ([xshift=-\pgflinewidth-1pt,yshift=-\pgflinewidth+1pt] queue2.north east)
    rectangle ([xshift=\pgflinewidth+4pt,yshift=\pgflinewidth-1pt] queue2.north west);
    \fill[white] ([xshift=-\pgflinewidth-1pt,yshift=-\pgflinewidth] queue2.north east)
    rectangle ([xshift=15pt,yshift=\pgflinewidth] queue2.north west);
    \draw[-] ([xshift=15pt] queue2.north east) -- ([xshift=15pt] queue2.north west);

    \node[serverps] (phi2) at ($(queue2.south)+(.55cm,0)$) {};
    \node at (phi2) {$\phi_2(x)$};
    \node[anchor=south] at ($(queue2)+(0,.35cm)$) {$x_2 = 2$};

    \draw[arrow,<->] ($(queue1.north)+(-.2cm,-.1cm)$) --
    node[midway,align=center,anchor=south,yshift=.08cm] {Class-$1$ \\ tokens}
    ($(nu)+(.9cm,.35cm)$);
    \draw[arrow,<->] ($(queue2.north)+(-.2cm,.1cm)$) --
    node[midway,align=center,anchor=north,yshift=-.08cm] {Class-$2$ \\ tokens}
    ($(nu)+(.9cm,-.35cm)$);

  \end{tikzpicture}
  \caption{A closed queueing system, consisting
  of an \gls{oi} queue and a Whittle network,
  associated with the server pool of \figurename~\ref{fig:tripartite}.
  At most $\ell_1 = \ell_2 = 4$ jobs can be assigned to each class.}
  \label{fig:oi+whittle}
\end{figure}
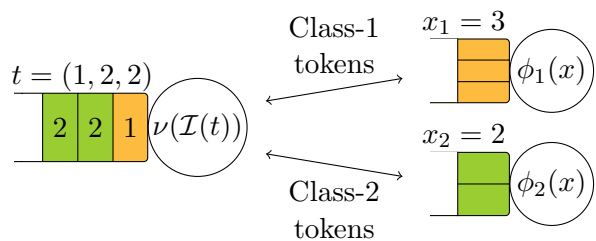

As announced earlier,
we can also combine our token-based load balancing algorithm
with balanced fairness.
The assignment of jobs to classes is still regulated
by a single bucket containing available tokens, sorted in release order,
but the resources are now allocated according to balanced fairness.
The corresponding queueing model
consists of an \gls{oi} queue and a Whittle network,
as represented in \figurename~\ref{fig:oi+whittle}.

The intermediary state $(x,t)$,
retaining the release order of available tokens but not the arrival order of jobs,
defines a Markov process.
Its stationary distribution follows from known results
on networks of quasi-reversible queues \cite{kelly}:
\begin{equation*}
  \pi(x,t) = \frac1G \Phi(x) \Lambda(t),
  \quad \forall x \in \states,
  ~\forall t \in \seq:
  ~x + |t| = \ell.
\end{equation*}
We can show as before that the average per-class arrival rates, ignoring the release order of tokens,
are as defined by the dynamic load balancing of \S \ref{subsec:random}.

The insensitivity of balanced fairness to the job size distribution within each class is again preserved.
The proof of \cite{BP02} for Cox distributions extends directly.
Note that this does no imply that performance
is insensitive to the job size distribution \emph{within each type}.
Indeed, if two job types with different size distributions can be assigned to the same class,
then the distribution of the job sizes within this class may be correlated to the system state upon their arrival.
This point will be assessed by simulation in
Section \ref{sec:num}.

Observe that
our token-based mechanism
can be applied to balance the load between the queues of an arbitrary Whittle network,
as represented in \figurename~\ref{fig:oi+whittle},
independently of the system considered.
Examples or such systems are given in \cite{BJP04}.

\section{Numerical results}
\label{sec:num}

We finally consider two examples that give insights on the performance of our token-based algorithm.
We especially make a comparison with the static load balancing of Section \ref{sec:resource-allocation}
and assess the insensitivity to the job size distribution within each type.
We refer the reader to \cite{J16} for a large-scale analysis
in homogeneous pools with a single job type,
along with a comparison with other (non-insensitive) standard policies.

Performance metrics for Poisson arrival processes
and exponentially distributed sizes with unit mean
follow from \eqref{eq:stationary}.
By insensitivity,
these also give the performance
when job sizes within each class are i.i.d.,
as long as the traffic intensity is unchanged.
We resort to simulations to evaluate performance
when the job size distribution is type-dependent.

Performance is measured by the job blocking probability
and the resource occupancy.
For each $k = 1,\ldots,K$, we let
\begin{equation*}
  \beta_k = \frac1G
  \sum_{\substack{x \le \ell: \\ x_i = \ell_i,~\forall i \in \classes: k \in \types_i}}
  \Phi(x) \Lambda(\ell-x)
\end{equation*}
denote the probability that a job of type $k$ is blocked upon arrival.
The equality follows from PASTA property \cite{S99}.
Symmetrically, for each $s = 1,\ldots,S$, we let
\begin{equation*}
  \psi_s = \frac1G
  \sum_{\substack{x \le \ell: \\ x_i = 0,~\forall i \in \classes: s \in \servers_i}}
  \Phi(x) \Lambda(\ell-x)
\end{equation*}
denote the probability that server $s$ is idle.
These quantities are related by the conservation equation
\begin{equation}
  \label{eq:conservation}
  \sum_{k=1}^K \nu_k (1 - \beta_k)
  = \sum_{s=1}^S \mu_s (1 - \psi_s).
\end{equation}
We define respectively the average blocking probability
and the average resource occupancy by
\begin{equation*}
  \beta = \frac{ \sum_{k=1}^K \nu_k \beta_k }{ \sum_{k=1}^K \nu_k }
  \quad \text{and} \quad
  \eta = \frac{ \sum_{s=1}^S \mu_s (1 - \psi_s) }{ \sum_{s=1}^S \mu_s }.
\end{equation*}
There is a simple relation between $\beta$ and $\eta$.
Indeed, if we let $\rho = ( \sum_{k=1}^K \nu_k ) / ( \sum_{s=1}^S \mu_s )$ denote the total load in the system,
then we can rewrite \eqref{eq:conservation} as
$\rho (1 - \beta) = \eta$.

As expected,
minimizing the average blocking probability
is equivalent to maximizing the average resource occupancy.
It is however convenient to look at both metrics in parallel.
As we will see,
when the system is underloaded,
jobs are almost never blocked
and it is easier to describe the (almost linear) evolution of the resource occupancy.
On the contrary, when the system is overloaded,
resources tend to be maximally occupied
and it is more interesting to focus on the blocking probability.

Observe that any stable server pool satisfies the conservation equation \eqref{eq:conservation}.
In particular, the average blocking probability $\beta$ in a stable system
cannot be less than $1 - \frac1\rho$ when $\rho > 1$.
A similar argument applied to each job type imposes that
\begin{equation}
  \label{eq:conservation-types}
  \beta_k
  \ge \max\left( 0, 1 - \frac1{\nu_k} \sum_{s \in \bigcup_{i:k \in \types_i} \servers_i} \mu_s \right),
\end{equation}
for each $k = 1,\ldots,K$.

\subsection{A single job type}
\label{subsec:single}

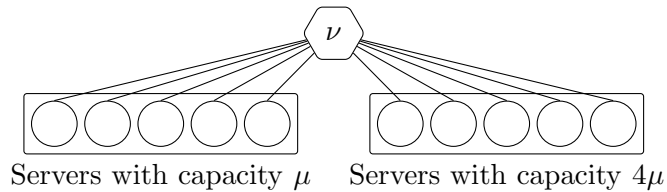
\begin{figure}[h]
  \centering
  \begin{tikzpicture}
    \def\width{.7cm}
    \def\height{1.2cm}

    \node[server, minimum size=.6cm] (server1) {};
    \node[server, minimum size=.6cm] (server2) at ($(server1)+(\width,0)$) {};
    \node[server, minimum size=.6cm] (server3) at ($(server2)+(\width,0)$) {};
    \node[server, minimum size=.6cm] (server4) at ($(server3)+(\width,0)$) {};
    \node[server, minimum size=.6cm] (server5) at ($(server4)+(\width,0)$) {};
    \node[server, minimum size=.6cm] (server6) at ($(server5)+(2.5*\width,0)$) {};
    \node[server, minimum size=.6cm] (server7) at ($(server6)+(\width,0)$) {};
    \node[server, minimum size=.6cm] (server8) at ($(server7)+(\width,0)$) {};
    \node[server, minimum size=.6cm] (server9) at ($(server8)+(\width,0)$) {};
    \node[server, minimum size=.6cm] (server10) at ($(server9)+(\width,0)$) {};

    \draw[rounded corners=.04cm] ($(server1)+(-.4cm,.4cm)$)
    rectangle node[midway,below,yshift=-.4cm] {Servers with capacity $\mu$}
    ($(server5)+(.4cm,-.4cm)$);
    \draw[rounded corners=.04cm] ($(server6)+(-.4cm,.4cm)$)
    rectangle node[midway,below,yshift=-.4cm] {Servers with capacity $4 \mu$}
    ($(server10)+(.4cm,-.4cm)$);

    \node[type] (type) at ($(server5)!.5!(server6)+(0,\height)$) {};
    \node at (type) {$\nu$};

    \foreach \a in {1,2,...,10} {
      \draw[-] (type) -- (server\a.north);
    }
  \end{tikzpicture}
  \caption{A server pool with a single job type.
  Classes are omitted because each of them corresponds to a single server.
  }
  \label{fig:single-graph}
\end{figure}

We first consider a pool of $S = 10$ servers
with a single type of jobs ($K = 1$),
as shown in \figurename~\ref{fig:single-graph}.
Each class identifies a unique server
and each job can be assigned to any class.
Half of the servers have a unit capacity $\mu$ and the other half have capacity $4 \mu$.
Each server has $\ell = 6$ tokens
and applies \gls{ps} policy to its jobs.
We do not look at the insensitivity to the job size distribution in this case,
as there is a single job type.

\paragraph*{Comparison}

\begin{figure}
  \centering
  \pgfplotstableread[col sep=comma]{theo-single-comp-hete.csv}\hete

\begin{tikzpicture}
\begin{axis}[
  defaultplots,
  legend pos=south east,
  xtick={0,.4,1,2,3,4},
  xticklabels={0,$\frac25$, 1, 2, 3, 4},
]

\addplot+[
  red, no markers, line width=.7pt,
] table[x=rho, y=db]{\hete};
\addlegendentry{Dynamic}

\addplot+[
  red, no markers, line width=.7pt, forget plot,
] table[x=rho, y=do]{\hete};

\addplot+[
  mygreenplot, dashdotted, no markers, line width=.7pt,
] table[x=rho, y=ob]{\hete};
\addlegendentry{Best static}

\addplot+[
  mygreenplot, dashdotted, no markers, line width=.7pt, forget plot,
] table[x=rho, y=oo]{\hete};

\addplot+[
  orange, densely dotted, no markers, line width=.7pt,
] table[x=rho, y=ub]{\hete};
\addlegendentry{Uniform static}

\addplot+[
  orange, densely dotted, no markers, line width=.7pt, forget plot,
] table[x=rho, y=uo]{\hete};

\addplot+[
  blue, densely dashed, no markers, line width=.7pt,
  domain=1:4, samples=100,
] {1 - 1/x};
\addlegendentry{Ideal}

\addplot+[
  blue, densely dashed, no markers, line width=.7pt, forget plot,
  domain=0:1, samples=2,
] {0};

\addplot+[
  blue, densely dashed, no markers, line width=.7pt, forget plot,
  domain=0:1, samples=100,
] {x};

\addplot+[
  blue, densely dashed, no markers, line width=.7pt, forget plot,
  domain=1:4, samples=2,
] {1};

\end{axis}
\end{tikzpicture}
  \caption{Performance of the dynamic load balancing in the pool of \figurename~\ref{fig:single-graph}.
  Average blocking probability (bottom plot) and resource occupancy (top plot).}
  \label{fig:single-comp}
\end{figure}
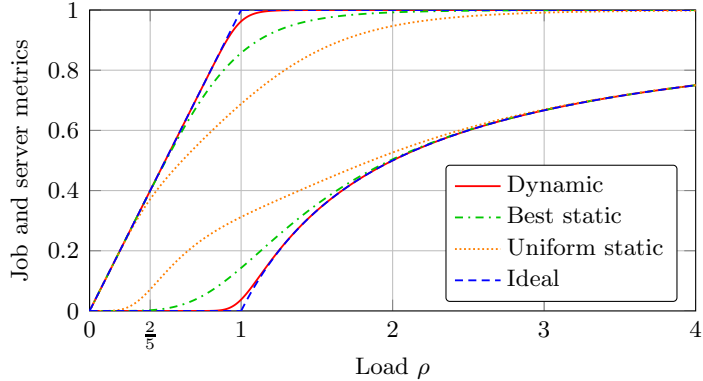

We compare the performance of our algorithm
with that of the static load balancing of Section \ref{sec:resource-allocation},
where each job is assigned to a server at random,
independently of system state,
and blocked if its assigned server is already full.
We consider two variants,
\textit{best static}~ and \textit{uniform static},
where the assignment probabilities are
proportional to the server capacities and uniform,
respectively.
\textit{Ideal} refers to the lowest average blocking probability that complies with the system stability.
According to \eqref{eq:conservation}, it is $0$ when $\rho \le 1$ and $1 - \frac1\rho$ when $\rho > 1$.
One can think of it as the performance in an ideal server pool
where resources would be constantly optimally utilized.
The results are shown in \figurename~\ref{fig:single-comp}.

The performance gain of our algorithm
compared to the static policies
is maximal near the critical load $\rho = 1$,
which is also the area where the delta with \textit{ideal} is maximal.
Elsewhere, all load balancing policies have a comparable performance.
Our intuition is as follows:
when the system is underloaded,
servers are often available and the blocking probability is low anyway;
when the system is overloaded,
resources are congested and the blocking probability is high whichever scheme is utilized.
Observe that the performance under \textit{uniform static}
deteriorates faster, even when $\rho < 1$,
because the servers with the lowest capacity,
concentrating half of the arrivals with only $\frac15$-th of the service capacity,
are congested whenever $\rho > \frac25$.
This stresses the need for accurate rate estimations
under a static load balancing.

\paragraph*{Asymptotics when the number of tokens increases}

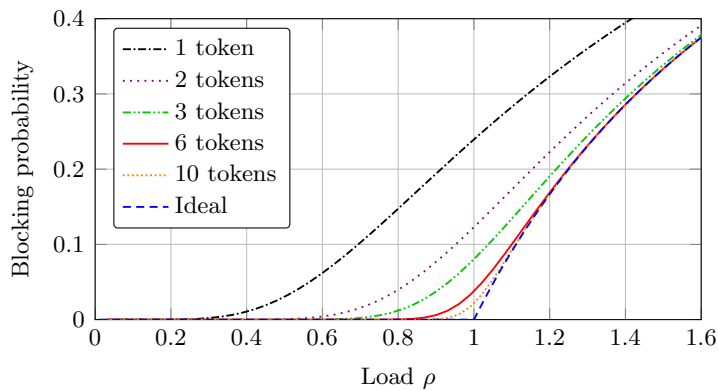
\begin{figure}[h]
  \centering
    \pgfplotstableread[col sep=comma]{theo-single-asym.csv}\t

  \begin{tikzpicture}
    \begin{axis}[
        defaultplots,
        xmin=0, xmax=1.6,
        ymin=0, ymax=.4,
        legend pos=north west,
        ylabel={Blocking probability},
        x label style = {yshift=-0.18cm},
      ]


      \addplot+[
        black, no markers, densely dashdotted, line width=.7pt,
      ] table[x=rho, y=1]{\t};
      \addlegendentry{$1$ token}

      \addplot+[
        violet, no markers, dotted, line width=.7pt,
      ] table[x=rho, y=2]{\t};
      \addlegendentry{$2$ tokens}

      \addplot+[
        mygreenplot, no markers, densely dashdotdotted, line width=.7pt,
      ] table[x=rho, y=3]{\t};
      \addlegendentry{$3$ tokens}

      \addplot+[
        red, no markers, solid, line width=.7pt,
      ] table[x=rho, y=6]{\t};
      \addlegendentry{$6$ tokens}

      \addplot+[
        orange, no markers, densely dotted, line width=.7pt,
      ] table[x=rho, y=10]{\t};
      \addlegendentry{$10$ tokens}


      \addplot+[
        blue, densely dashed, no markers, line width=.7pt, domain=.01:1.6, samples=400,
      ]{max(0,1 - 1 / x)};
      \addlegendentry{Ideal}

    \end{axis}
  \end{tikzpicture}
  \caption{Impact of the number of tokens
  on the average blocking probability
  under the dynamic load balancing
  in the pool of \figurename~\ref{fig:single-graph}.}
  \label{fig:single-asym}
\end{figure}

We now focus on the impact of the number of tokens
on the performance of the dynamic load balancing.
A direct calculation shows
that the average blocking probability decreases with the number $\ell$ of tokens per server,
and tends to \textit{ideal} as $\ell \to +\infty$.
Intuitively, having many tokens gives a long run feedback on the server loads
without blocking arrivals more than necessary (to preserve stability).
The results are shown in \figurename~\ref{fig:single-asym}.

We observe that the convergence to the asymptotic ideal is quite fast.
The largest gain is obtained with small values of $\ell$
and the performance is already close to the optimal with $\ell = 10$ tokens per server.
Hence, we can reach a low blocking probability
even when the number of tokens is limited,
for instance to guarantee a minimum service rate per job
or respect multitasking constraints on the servers.

\subsection{Several job types}
\label{subsec:multi}

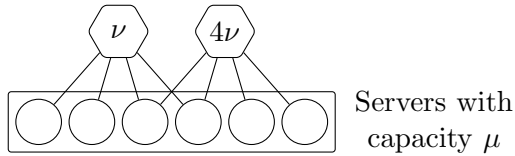
\begin{figure}[h]
  \centering
  \begin{tikzpicture}
    \def\width{.7cm}
    \def\height{1.2cm}

    \node[server,minimum size=.6cm] (server1) {};
    \node[server,minimum size=.6cm] (server2) at ($(server1)+(\width,0)$) {};
    \node[server,minimum size=.6cm] (server3) at ($(server2)+(\width,0)$) {};
    \node[server,minimum size=.6cm] (server4) at ($(server3)+(\width,0)$) {};
    \node[server,minimum size=.6cm] (server5) at ($(server4)+(\width,0)$) {};
    \node[server,minimum size=.6cm] (server6) at ($(server5)+(\width,0)$) {};

    \node[type] (type1) at ($(server2)!.5!(server3)+(0,\height)$) {};
    \node[type] (type2) at ($(server4)!.5!(server5)+(0,\height)$) {};
    \node at (type1) {$\nu$};
    \node at (type2) {$4 \nu$};

    \draw[-] (type1) -- (server1);
    \draw[-] (type1) -- (server2);
    \draw[-] (type1) -- (server3);
    \draw[-] (type1) -- (server4);
    \draw[-] (type2) -- (server3);
    \draw[-] (type2) -- (server4);
    \draw[-] (type2) -- (server5);
    \draw[-] (type2) -- (server6);

    \draw[rounded corners=.04cm] ($(server1)+(-.4cm,.4cm)$)
    rectangle node[near end, align=center, xshift=1.2cm, yshift=.2cm, anchor=west] {Servers with \\ capacity $\mu$}
    ($(server6)+(.4cm,-.4cm)$);
  \end{tikzpicture}
  \caption{A server pool with two job types.}
  \label{fig:multi-graph}
\end{figure}

We now consider a pool of $S = 6$ servers,
all with the same unit capacity $\mu$,
as shown in \figurename~\ref{fig:multi-graph}.
As before,
there is no parallel processing.
Each class identifies a unique server
that applies \gls{ps} policy to its jobs
and has $\ell = 6$ tokens.
There are two job types with different arrival rates and compatibilities.
Type-$1$ jobs have a unit arrival rate $\nu$
and can be assigned to any of the first four servers.
Type-$2$ jobs arrive at rate $4 \nu$
and can be assigned to any of the last four servers.
Thus only two servers can be accessed by both types.
Note that heterogeneity now lies in the job arrival rates
and not in the server capacities.

\paragraph*{Comparison}

\begin{figure}
  \centering
    \pgfplotstableread[col sep=comma]{theo-multi-comp-hete.csv}\hete
  \begin{tikzpicture}
    \begin{axis}[
        defaultplots,
        legend pos=south east,
        xtick={0, .83333, 1, 1.66667, 2, 3, 4},
        xticklabels={0, $\frac56$, 1, $\frac53$, 2, 3, 4},
      ]

      \addplot+[
        red, no markers, solid, line width=.7pt,
      ] table[x=rho, y=do]{\hete};
      \addlegendentry{Dynamic}

      \addplot+[
        red, no markers, solid, line width=.7pt, forget plot,
      ] table[x=rho, y=db]{\hete};

      \addplot+[
        mygreenplot, no markers, dashdotted, line width=.7pt,
      ] table[x=rho, y=oo]{\hete};
      \addlegendentry{Best static}

      \addplot+[
        mygreenplot, no markers, dashdotted, line width=.7pt, forget plot,
      ] table[x=rho, y=ob]{\hete};

      \addplot+[
        orange, no markers, densely dotted, line width=.7pt,
      ] table[x=rho, y=uo]{\hete};
      \addlegendentry{Uniform static}

      \addplot+[
        orange, no markers, densely dotted, line width=.7pt, forget plot,
      ] table[x=rho, y=ub]{\hete};

      \addplot+[
        blue, no markers, densely dashed, line width=.7pt,
        domain=0:5/6, samples=100,
      ] {x};
      \addlegendentry{Ideal}

      \addplot+[
        blue, no markers, densely dashed, line width=.7pt, forget plot,
        domain=5/6:5/3, samples=100,
      ] {5/6 + (x-5/6)/5};

      \addplot+[
        blue, no markers, densely dashed, line width=.7pt, forget plot,
        domain=5/3:4, samples=100,
      ] {1};

      \addplot+[
        blue, no markers, densely dashed, line width=.7pt, forget plot,
        domain=0:5/6, samples=100,
      ] {0};

      \addplot+[
        blue, no markers, densely dashed, line width=.7pt, forget plot,
        domain=5/6:5/3, samples=100,
      ] {(4/5)*(1 - 5/(6*x))};

      \addplot+[
        blue, no markers, densely dashed, line width=.7pt, forget plot,
        domain=5/3:4, samples=100,
      ] {1 - 1/x};
    \end{axis}
  \end{tikzpicture}
  \caption{Performance of the dynamic load balancing
  in the pool of \figurename~\ref{fig:multi-graph}.
  Average blocking probability (bottom plot) and resource occupancy (top plot).}
  \label{fig:multi-comp}
\end{figure}
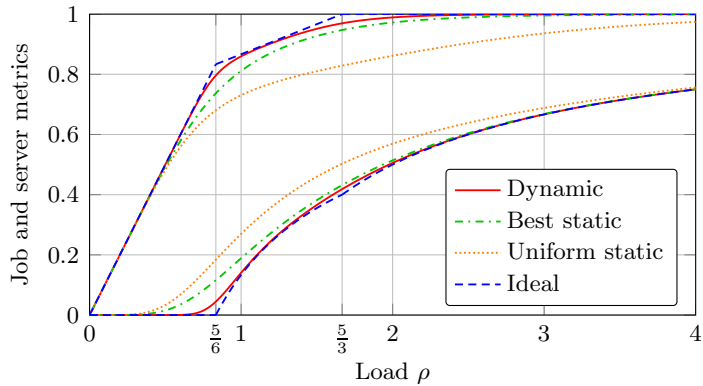

We again consider
two variants of the static load balancing:
\textit{best static},
in which the assignment probabilities are
chosen so as to homogenize the arrival rates at the servers as far as possible,
and \textit{uniform static},
in which the assignment probabilities are uniform.
Note that \textit{best static}
assumes that the arrival rates of the job types are known,
while \textit{uniform static} does not.
As before,
\textit{ideal} refers to the lowest average blocking probability
that complies with the system stability.
The results are shown in \figurename~\ref{fig:multi-comp}.

Regardless of the policy,
the slope of the resource occupancy breaks down
near the critical load $\rho = \frac56$.
The reason is that
the last four servers support at least $\frac45$-th of the arrivals
with only $\frac23$-rd of the service capacity,
so that their effective load is $\frac65 \rho$.
It follows from \eqref{eq:conservation-types} that
the average blocking probability in a stable system
cannot be less than $\frac45 (1 - \frac56 \frac1\rho)$
when $\rho \ge \frac56$.
Under \textit{ideal},
the slope of the resource occupancy breaks down again at $\rho = \frac53$.
This is the point where
the first two servers cannot support the load of type-$1$ jobs by themselves anymore.

Otherwise, most of the observations of \S \ref{subsec:single} are still valid.
The performance gain of the dynamic load balancing
compared to \textit{best static}
is maximal near the first critical load $\rho = \frac56$.
Its delta with \textit{ideal}
is maximal near $\rho = \frac56$ and $\rho = \frac53$.
Elsewhere,
all schemes have a similar performance,
except for \textit{uniform static} that deteriorates faster.

Overall, these numerical results show that
our dynamic load balancing algorithm often outperforms
\textit{best static}
and is close to \textit{ideal}.
The configurations (not shown here) where
it was not the case
involved very small pools,
with job arrival rates and compatibilities opposite to the server capacities.
Our intuition is that our algorithm performs better
when the pool size or the number of tokens
allow for some diversity in the assignments.

\paragraph*{(In)sensitivity}

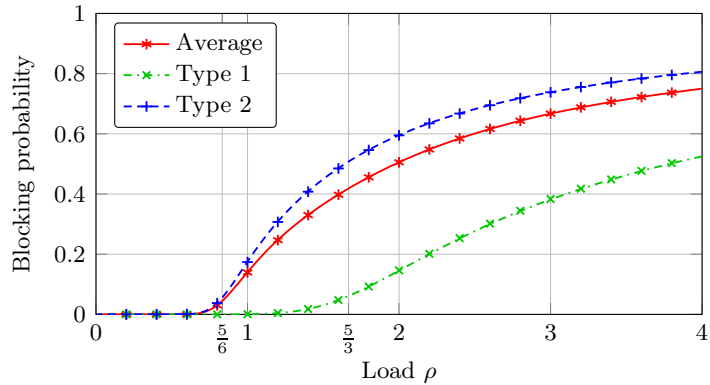
\begin{figure}[h]
  \centering
    \pgfplotstableread[col sep=comma]{theo-multi-comp-hete.csv}\hete
  \pgfplotstableread[col sep=comma]{multi-hete-dynamic-correlated-hyperexp-interval-blocking.csv}\simu

  \begin{tikzpicture}
    \begin{axis}[
        defaultplots,
        legend pos=north west,
        ylabel={Blocking probability},
        xtick={0, .83333, 1, 1.66667, 2, 3, 4},
        xticklabels={0, $\frac56$, 1, $\frac53$, 2, 3, 4},
      ]

      \addlegendimage{red, mark=asterisk, mark options=solid, solid, line width=.7pt}
      \addlegendentry{Average}

      \addlegendimage{mygreenplot, mark=x, mark options=solid, dashdotted, line width=.7pt}
      \addlegendentry{Type 1}

      \addlegendimage{blue, mark=+, mark options=solid, densely dashed, line width=.7pt}
      \addlegendentry{Type 2}


      \addplot+[
        red, no markers, solid, line width=.7pt, forget plot,
      ] table[x=rho, y=db]{\hete};

      \addplot+[
        red, only marks, mark=asterisk, line width=.7pt, forget plot,
      ] table[y index=5]{\simu};


      \addplot+[
        mygreenplot, no markers, dashdotted, line width=.7pt, forget plot,
      ] table[x=rho, y=db2]{\hete};

      \addplot+[
        mygreenplot, only marks, mark=x, line width=.7pt, forget plot,
      ] table[y index=3]{\simu};


      \addplot+[
        blue, no markers, densely dashed, line width=.7pt, forget plot,
      ] table[x=rho, y=db1]{\hete};

      \addplot+[
        blue, only marks, mark=+, line width=.7pt, forget plot,
      ] table[y index=1]{\simu};

    \end{axis}
  \end{tikzpicture}
  \caption{Blocking probability under the dynamic load balancing
    in the server pool of \figurename~\ref{fig:multi-graph},
    with either exponentially distributed job sizes (line plots)
    or hyperexponentially distributed sizes (marks).
    Each simulation point is the average of 100 independent runs,
    each built up of $10^6$ jumps after a warm-up period of $10^6$ jumps.
    The corresponding $95\%$ confidence interval, not shown on the figure,
    does not exceed $\pm0.001$ around the point.
  }
  \label{fig:multi-sens}
\end{figure}

We finally evaluate the sensitivity of our algorithm to the job size distribution within each type.
\figurename~\ref{fig:multi-sens} shows the results.
Lines give the performance
when job sizes are exponentially distributed with unit mean, as before.
Marks, obtained by simulation,
give the performance
when the job size distribution within each type is hyperexponential:
$\frac13$-rd of type-$1$ jobs have an exponentially distributed size with mean $2$
and the other $\frac23$-rd have an exponentially distributed size with mean $\frac12$;
similarly, $\frac16$-th of type-$2$ jobs have an exponentially distributed size with mean $5$
and the other $\frac56$-th have an exponentially distributed size with mean $\frac15$.

The similarity of the exact and simulation results suggests that
insensitivity is preserved even when the job size distribution is type-dependent.
Further evaluations, involving other job size distributions,
would be necessary to conclude.

Also observe that the blocking probability of type-$1$ jobs
increases near the load $\rho = \frac53$,
which is twice less than
the upper bound $\rho = \frac{10}3$ given by \eqref{eq:conservation-types}.
This suggests that the dynamic load balancing
compensates the overload of type-$2$ jobs by rejecting more jobs of type $1$.

\section{Conclusion}
\label{sec:ccl}

We have introduced a new server pool model
that explicitly distinguishes
the compatibilities of a job from its actual assignment by the load balancer.
Expressing the results of \cite{BJP04} in this new model
has allowed us to see the problem of load balancing in a new light.
We have derived a deterministic, token-based implementation of a dynamic load balancing
that preserves the insensitivity of balanced fairness to the job size distribution within each class.
Numerical results have assessed the performance of this algorithm.

For the future works,
we would like to evaluate the performance of our algorithm
in broader classes of server pools.
We are also interested in proving its insensitivity
to the job size distribution within each type.


\begin{thebibliography}{10}

\bibitem{alis}
I.~Adan and G.~Weiss.
\newblock A loss system with skill-based servers under assign to longest idle
  server policy.
\newblock {\em Probability in the Engineering and Informational Sciences},
  26(3):307–321, 2012.

\bibitem{B13}
L.~A. Barroso, J.~Clidaras, and U.~H\"olzle.
\newblock {\em The Datacenter As a Computer: An Introduction to the Design of
  Warehouse-Scale Machines}.
\newblock Morgan \& Claypool Publishers, 2nd edition, 2013.

\bibitem{BK96}
S.~A. Berezner and A.~E. Krzesinski.
\newblock Order independent loss queues.
\newblock {\em Queueing Systems}, 23(1-4):331--335, Mar. 1996.

\bibitem{BC17}
T.~Bonald and C.~Comte.
\newblock Balanced fair resource sharing in computer clusters.
\newblock {\em Performance Evaluation}, 116(Supplement C):70--83, Nov. 2017.

\bibitem{BJP04}
T.~Bonald, M.~Jonckheere, and A.~Prouti{\`e}re.
\newblock Insensitive load balancing.
\newblock {\em SIGMETRICS Perform. Eval. Rev.}, 32(1):367--377, June 2004.

\bibitem{BP02}
T.~Bonald and A.~Prouti{\`e}re.
\newblock Insensitivity in processor-sharing networks.
\newblock {\em Performance Evaluation}, 49(1):193--209, Sept. 2002.

\bibitem{BP03-1}
T.~Bonald and A.~Prouti{\`e}re.
\newblock Insensitive bandwidth sharing in data networks.
\newblock {\em Queueing Syst.}, 44(1):69--100, 2003.

\bibitem{F05}
S.~Fujishige.
\newblock {\em Submodular {Functions} and {Optimization}, {Volume} 58 - 2nd
  {Edition}}.
\newblock Elsevier Science, July 2005.

\bibitem{G15}
K.~Gardner, S.~Zbarsky, S.~Doroudi, M.~Harchol-Balter, and E.~Hyytia.
\newblock Reducing latency via redundant requests: Exact analysis.
\newblock {\em SIGMETRICS Perform. Eval. Rev.}, 43(1):347--360, June 2015.

\bibitem{J16}
M.~Jonckheere and B.~J. Prabhu.
\newblock Asymptotics of insensitive load balancing and blocking phases.
\newblock {\em SIGMETRICS Perform. Eval. Rev.}, 44(1):311--322, June 2016.

\bibitem{kelly}
F.~P. Kelly.
\newblock {\em Reversibility and {Stochastic} {Networks}}.
\newblock Cambridge University Press, New York, NY, USA, 2011.

\bibitem{K11}
A.~E. Krzesinski.
\newblock Order independent queues.
\newblock In R.~J. Boucherie and N.~M. van Dijk, editors, {\em Queueing
  Networks: A Fundamental Approach}, pages 85--120. Springer US, Boston, MA,
  2011.

\bibitem{jiq}
Y.~Lu, Q.~Xie, G.~Kliot, A.~Geller, J.~R. Larus, and A.~Greenberg.
\newblock Join-{Idle}-{Queue}: {A} novel load balancing algorithm for
  dynamically scalable web services.
\newblock {\em Performance Evaluation}, 68(11):1056--1071, Nov. 2011.

\bibitem{S99}
R.~Serfozo.
\newblock {\em Introduction to Stochastic Networks}.
\newblock Stochastic Modelling and Applied Probability. Springer New York,
  1999.

\bibitem{SV15}
V.~Shah and G.~de~Veciana.
\newblock High-{Performance} {Centralized} {Content} {Delivery}
  {Infrastructure}: {Models} and {Asymptotics}.
\newblock {\em IEEE/ACM Transactions on Networking}, 23(5):1674--1687, Oct.
  2015.

\bibitem{BBL17}
M.~{van der Boor}, S.~Borst, and J.~{van Leeuwaarden}.
\newblock Load balancing in large-scale systems with multiple dispatchers.
\newblock In {\em Proceedings of INFOCOM 2017}, 2017.

\end{thebibliography}

\appendix

\section*{Appendix: Proof of the irreducibility}

We prove the irreducibility of the Markov process defined by the state $(c,t)$
of a tandem network of two \gls{oi} queues, as described in \S \ref{subsec:deterministic}.
Throughout the proof,
we will simply refer to such a network as a \emph{tandem network},
implicitly meaning that it is as described in \S \ref{subsec:deterministic}.

\paragraph{Assumptions}

We first recall and name the two main assumptions that we use in the proof.
\begin{itemize}
  \item \emph{Positive service rate}.
    For each $i \in \classes$,
    $\types_i \neq \emptyset$ and $\servers_i \neq \emptyset$.
  \item \emph{Separability}.
    For each pair $\{i,j\} \subset \classes$,
    either $\servers_i \neq \servers_j$ or $\types_i \neq \types_j$ (or both).
\end{itemize}

\paragraph{Result statement}
The Markov process defined by the state of the tandem network is irreducible
on the state space $\espace = \{ (c,t) \in \seq^2: |c| + |t| = \ell \}$
comprising all states
with $\ell_i$ tokens of class $i$, for each $i \in \classes$.

\paragraph{Outline of the proof}

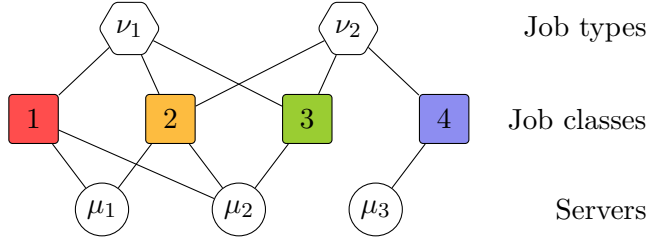
\begin{figure}
  \centering
  \begin{tikzpicture}
    \def\width{1.8cm}
    \def\height{1.2cm}

    \node[class,fill=myred!70] (class1) {1};
    \node[class,fill=myorange] (class2) at ($(class1)+(\width,0)$) {2};
    \node[class,fill=mygreen] (class3) at ($(class2)+(\width,0)$) {3};
    \node[class,fill=myblue!70] (class4) at ($(class3)+(\width,0)$) {4};

    \node[server] (server1) at ($(class1)!.5!(class2)-(0,\height)$) {};
    \node[server] (server2) at ($(class2)!.5!(class3)-(0,\height)$) {};
    \node[server] (server3) at ($(class3)!.5!(class4)-(0,\height)$) {};
    \node at (server1) {$\mu_1$};
    \node at (server2) {$\mu_2$};
    \node at (server3) {$\mu_3$};

    \node[type] (type1) at ($(class1)!.7!(class2)+(0,\height)$) {};
    \node[type] (type2) at ($(class4)!.7!(class3)+(0,\height)$) {};
    \node at (type1) {$\nu_1$};
    \node at (type2) {$\nu_2$};

    \draw (type1) -- (class1) (type1) -- (class2) (type1) -- (class3);
    \draw (type2) -- (class2) (type2) -- (class3) (type2) -- (class4);
    \draw (server1) -- (class1) (server1) -- (class2);
    \draw (server2) -- (class1) (server2) -- (class2) (server2) -- (class3);
    \draw (server3) -- (class4);

    \node[anchor=east] at ($(class4)+(2.8cm,0)+(0,\height)$) {Job types};
    \node[anchor=east] at ($(class4)+(2.8cm,0)$) {Job classes};
    \node[anchor=east] at ($(class4)+(2.8cm,0)-(0,\height)$) {Servers};
  \end{tikzpicture}
  \caption{
  A technically interesting toy configuration.
  We have $\types_2 = \types_3$ and $\servers_3 \subsetneq \servers_2$,
  so that class-$2$ tokens can overtake class-$3$ tokens in the queue of tokens held by jobs in service
  but not in the queue of available tokens.
  On the other hand, $\types_1 \subsetneq \types_2$ and $\servers_1 = \servers_2$,
  so that class-$2$ tokens can overtake class-$1$ tokens in the queue of available tokens
  but not in the queue of tokens held by jobs in service.
  In none of the queues can class-$2$ tokens
  overtake tokens of classes $1$ and $3$ at once.
  }
  \label{fig:tripartite-app}
\end{figure}

We provide a constructive proof
that exhibits a series of transitions
leading from any feasible state to any feasible state with a nonzero probability.
We first describe two types of transitions
and specify the states where they can occur with a nonzero probability.
\begin{itemize}
  \item \emph{Circular shift}:
    service completion of a token at the head of a queue.
    This transition
    is always possible
    thanks to the positive service rate assumption.
    Consequently,
    states that are circular shifts of each other can communicate.
    We will therefore focus on ordering tokens relative to each other,
    keeping in mind that we can eventually apply circular shifts
    to move them in the correct queue.
  \item \emph{Overtaking}:
    service completion of a token that is in second position of a queue,
    before its predecessor completes service.
    Such a transition has the effect of
    swapping the order of these two tokens.
    By reindexing classes if necessary,
    we can work on the assumption that class-$i$ tokens can overtake the tokens of classes $1$ to $i-1$
    in (at least) one of the two queues,
    for each $i = 2,\ldots,N$.
    The proof of this statement relies on the separability assumption.
\end{itemize}
It is tempting to consider more sophisticated transitions,
for instance where a token overtakes several other tokens at once.
Unfortunately, our assumptions do not guarantee that such transitions can occur with a nonzero probability.
An example is shown in \figurename~\ref{fig:tripartite-app}.
The two operations \emph{circular shift} and \emph{overtaking}
will prove to be sufficient.
We first combine them to show the following intermediary result:
\begin{itemize}
  \item From any feasible state,
    we can reach the state where all class-$N$ tokens are gathered at some selected position in one of the two queues
    while the position of the other tokens is unchanged.
\end{itemize}
We finally prove the irreducibility result by induction on the number $N$ of classes.
As announced,
the proof is constructive:
it gives a series of transitions
leading from any state to any other state.
The induction step can be decomposed in two parts:
\begin{itemize}
  \item By repeatedly moving class-$N$ tokens at a position where they do not prevent other tokens from overtaking each other,
    we can order the tokens of classes $1$ to $N-1$ as if class-$N$ tokens were absent.
    The induction assumption ensures that we can perform this reordering.
  \item Once the tokens of classes $1$ to $N-1$ are well ordered, class-$N$ tokens can be positioned among them.
\end{itemize}
We now detail the steps of the proof one after the other.

\paragraph{Circular shift}

Because of the positive service rate assumption,
a token at the head of either of the two queues has a nonzero probability of
completing service and moving to the end of the other queue.
We refer to such a transition as a \emph{circular shift}.

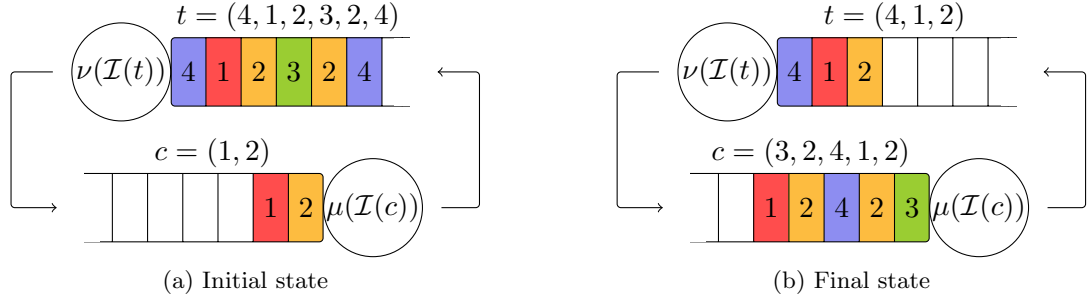
\begin{figure}[h]
  \centering
  \subfloat[Initial state\label{fig:initial}]{
  \begin{tikzpicture}
    \node[fcfs=8,
      rectangle split part fill={white,white,white,white,white,white,myred!70,myorange},
    ] (queue) {
      \nodepart {one} {}
      \nodepart {two} {}
      \nodepart {three} {\phantom{2}}
      \nodepart {four} {\phantom{2}}
      \nodepart {five} {\phantom{2}}
      \nodepart {six} {\phantom{2}}
      \nodepart {seven} {1}
      \nodepart {eight} {2}
    };
    \fill[white] ([xshift=-\pgflinewidth-1pt,yshift=-\pgflinewidth+1pt]queue.north west)
    rectangle ([xshift=\pgflinewidth+4pt,yshift=\pgflinewidth-1pt]queue.south west);
    \fill[white] ([xshift=-\pgflinewidth-1pt,yshift=-\pgflinewidth-.045pt]queue.north west)
    rectangle ([xshift=\pgflinewidth+10pt,yshift=\pgflinewidth+.045pt]queue.south west);
    \node[anchor=south,align=center] at ($(queue)+(.2cm,.4cm)$) {$c = (1,2)$};

    \node[serveroi] (mu) at ($(queue.east)+(.65cm,0)$) {};
    \node at (mu) {$\mu(\classes(c))$};

    \node[fcfs=8,
      rectangle split part fill={myblue!70,myred!70,myorange,mygreen,myorange,myblue!70,white},
    ] (bucket) at ($(queue)+(1.3cm,1.8cm)$) {
      \nodepart {one} {4}
      \nodepart {two} {1}
      \nodepart {three} {2}
      \nodepart {four} {3}
      \nodepart {five} {2}
      \nodepart {six} {4}
      \nodepart {seven} {}
      \nodepart {eight} {}
    };
    \fill[white] ([xshift=\pgflinewidth+1pt,yshift=-\pgflinewidth+1pt]bucket.north east)
    rectangle ([xshift=-\pgflinewidth-4pt,yshift=\pgflinewidth-1pt]bucket.south east);
    \fill[white] ([xshift=\pgflinewidth+1pt,yshift=-\pgflinewidth-.045pt]bucket.north east)
    rectangle ([xshift=-\pgflinewidth-10pt,yshift=\pgflinewidth+.045pt]bucket.south east);
    \node[anchor=south,align=center] at ($(bucket)+(-.1cm,.4cm)$) {$t = (4,1,2,3,2,4)$};

    \node[serveroi] (nu) at ($(bucket.west)-(.65cm,0)$) {};
    \node at (nu) {$\nu(\classes(t))$};

    \draw[arrow] ($(mu)+(.9cm,0)$) -- ($(mu)+(1.4cm,0)$) |- ($(bucket.east)+(.2cm,0)$);
    \draw[arrow] ($(nu)-(.9cm,0)$) -| ($(queue.west)+(-.8cm,0)$) -- ($(queue.west)+(-.2cm,0)$);
  \end{tikzpicture}
  }
  \qquad \qquad
  \subfloat[Final state\label{fig:final}]{
  \begin{tikzpicture}
    \node[fcfs=8,
      rectangle split part fill={white,white,white,myred!70,myorange,myblue!70,myorange,mygreen},
    ] (queue) {
      \nodepart {one} {}
      \nodepart {two} {}
      \nodepart {three} {\phantom{2}}
      \nodepart {four} {1}
      \nodepart {five} {2}
      \nodepart {six} {4}
      \nodepart {seven} {2}
      \nodepart {eight} {3}
    };
    \fill[white] ([xshift=-\pgflinewidth-1pt,yshift=-\pgflinewidth+1pt]queue.north west)
    rectangle ([xshift=\pgflinewidth+4pt,yshift=\pgflinewidth-1pt]queue.south west);
    \fill[white] ([xshift=-\pgflinewidth-1pt,yshift=-\pgflinewidth-.045pt]queue.north west)
    rectangle ([xshift=\pgflinewidth+10pt,yshift=\pgflinewidth+.045pt]queue.south west);
    \node[anchor=south,align=center] at ($(queue)+(.1cm,.4cm)$) {$c = (3,2,4,1,2)$};

    \node[serveroi] (mu) at ($(queue.east)+(.65cm,0)$) {};
    \node at (mu) {$\mu(\classes(c))$};

    \node[fcfs=8,
      rectangle split part fill={myblue!70,myred!70,myorange,white},
    ] (bucket) at ($(queue)+(1.3cm,1.8cm)$) {
      \nodepart {one} {4}
      \nodepart {two} {1}
      \nodepart {three} {2}
      \nodepart {four} {\phantom{2}}
      \nodepart {five} {\phantom{2}}
      \nodepart {six} {\phantom{2}}
      \nodepart {seven} {}
      \nodepart {eight} {}
    };
    \fill[white] ([xshift=\pgflinewidth+1pt,yshift=-\pgflinewidth+1pt]bucket.north east)
    rectangle ([xshift=-\pgflinewidth-4pt,yshift=\pgflinewidth-1pt]bucket.south east);
    \fill[white] ([xshift=\pgflinewidth+1pt,yshift=-\pgflinewidth-.045pt]bucket.north east)
    rectangle ([xshift=-\pgflinewidth-10pt,yshift=\pgflinewidth+.045pt]bucket.south east);
    \node[anchor=south,align=center] at ($(bucket)+(-.1cm,.4cm)$) {$t = (4,1,2)$};

    \node[serveroi] (nu) at ($(bucket.west)-(.65cm,0)$) {};
    \node at (nu) {$\nu(\classes(t))$};

    \draw[arrow] ($(mu)+(.9cm,0)$) -- ($(mu)+(1.4cm,0)$) |- ($(bucket.east)+(.2cm,0)$);
    \draw[arrow] ($(nu)-(.9cm,0)$) -| ($(queue.west)+(-.8cm,0)$) -- ($(queue.west)+(-.2cm,0)$);
  \end{tikzpicture}
  }
  \caption{Circular shift.
  Sequence of transitions to reach state (b) from state (a):
  all tokens complete service in the first queue;
  all tokens before that of class $3$ complete service in the second queue;
  the first two tokens complete service in the first queue.
  }
  \label{fig:permutation}
\end{figure}

Now let $(c,t) \in \espace$ and $(c^\prime,t^\prime) \in \espace$,
with $c = (c_1,\ldots,c_n)$, $t = (t_1,\ldots,t_m)$,
$c^\prime = (c^\prime_1,\ldots,c^\prime_{n^\prime})$ and $t^\prime = (t^\prime_1,\ldots,t^\prime_{m^\prime})$.
Assume that the sequence $(c_1,\ldots,c_n,t_1,\ldots,t_m)$
is a circular shift
of the sequence $(c^\prime_1,\ldots,c^\prime_{n^\prime},t^\prime_1,\ldots,t^\prime_{m^\prime})$.
Then we can reach state $(c^\prime,t^\prime)$ from state $(c,t)$
by applying many circular shifts if necessary.
An example is shown in \figurename~\ref{fig:permutation} for the configuration of \figurename~\ref{fig:tripartite-app}.
All states that are circular shifts of each other can therefore communicate.

\paragraph{Overtaking}

We say that a token in second position of one of the two queues
\emph{overtakes} its predecessor
if it completes service first.
Such a transition allows us to exchange
the positions of these two tokens,
therefore escaping circular shifts to access other states.

Can such a transition occur with a nonzero probability?
It depends
on the classes of the tokens
in second and first positions,
denoted by $i$ and $j$ respectively.
The token in second position
can overtake its predecessor
if it receives a nonzero service rate.
In the queue of tokens held by jobs in service,
this means that there is at least one server that can process
class-$i$ jobs but not class-$j$ jobs,
that is $\servers_i \nsubseteq \servers_j$.
In the queue of available tokens,
this means that there is at least one job type that can seize
class-$i$ tokens but not class-$j$ tokens,
that is $\types_i \nsubseteq \types_j$.
Since states that are circular shifts of each other can communicate,
the queue where the overtaking actually occurs does not matter.

The separability assumption ensures that,
for each pair of classes,
the tokens of at least one of the two classes can overtake
the tokens of the other class, in at least one of the two queues.
We now show a stronger result:
by reindexing classes if necessary,
we can work on the assumption that
class-$i$ tokens can overtake the tokens of classes $1$ to $i-1$ in at least one of the two queues (possibly not the same),
for each $i = 2,\ldots,N$.

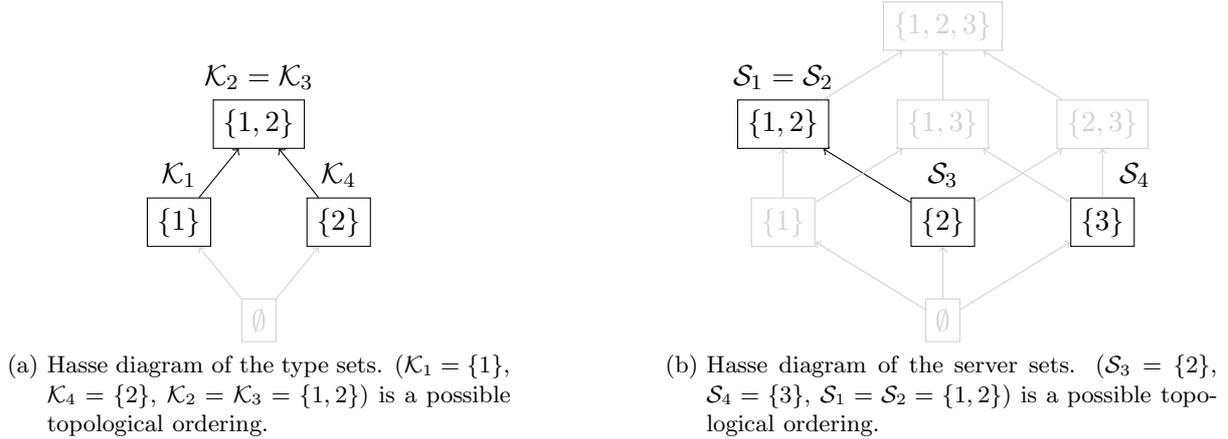
\begin{figure}[h]
  \centering
  \subfloat[Hasse diagram of the type sets.
  $(\types_1 = \{1\}$, $\types_4 = \{2\}$, $\types_2 = \types_3 = \{1,2\})$ is a possible topological ordering.
  \label{fig:typesets}
  ]{
  \makebox[.4\textwidth]{
    \begin{tikzpicture}
      \def\width{2.1cm}
      \def\height{1.3cm}

      \node[draw=mygray,text=mygray] (empty) {$\emptyset$};

      \node[draw] (1) at ($(empty)-(.5*\width,0)+(0,\height)$) {$\{1\}$};
      \node[anchor=south] at (1.north) {$\types_1$};
      \node[draw] (2) at ($(empty)+(.5*\width,0)+(0,\height)$) {$\{2\}$};
      \node[anchor=south] at (2.north) {$\types_4$};

      \node[draw] (12) at ($(1)!.5!(2)+(0,\height)$) {$\{1,2\}$};
      \node[anchor=south] at (12.north) {$\types_2 = \types_3$};

      \draw[->,mygray] (empty) -- (1);
      \draw[->,mygray] (empty) -- (2);
      \draw[->] (1) -- (12);
      \draw[->] (2) -- (12);
    \end{tikzpicture}
    }
  }
  \hfill
  \subfloat[Hasse diagram of the server sets.
  $(\servers_3 = \{2\}$, $\servers_4 = \{3\}$, $\servers_1 = \servers_2 = \{1,2\})$ is a possible topological ordering.
  \label{fig:serversets}
  ]{
  \makebox[.44\textwidth]{
    \begin{tikzpicture}
      \def\width{2.1cm}
      \def\height{1.3cm}

      \node[draw,mygray] (empty) {$\emptyset$};

      \node[draw,mygray] (1) at ($(empty)-(\width,0)+(0,\height)$) {$\{1\}$};
      \node[draw] (2) at ($(empty)+(0,\height)$) {$\{2\}$};
      \node[anchor=south] at (2.north) {$\servers_3$};
      \node[draw] (3) at ($(empty)+(\width,0)+(0,\height)$) {$\{3\}$};
      \node[anchor=south] at (3.north east) {$\servers_4$};

      \node[draw] (12) at ($(1)+(0,\height)$) {$\{1,2\}$};
      \node[anchor=south] at (12.north) {$\servers_1 = \servers_2$};
      \node[draw,mygray] (13) at ($(2)+(0,\height)$) {$\{1,3\}$};
      \node[draw,mygray] (23) at ($(3)+(0,\height)$) {$\{2,3\}$};

      \node[draw,mygray] (123) at ($(13)+(0,\height)$) {$\{1,2,3\}$};

      \draw[->,mygray] (empty) -- (1);
      \draw[->,mygray] (empty) -- (2);
      \draw[->,mygray] (empty) -- (3);
      \draw[->,mygray] (1) -- (12);
      \draw[->,mygray] (12) -- (123);
      \draw[->] (2) -- (12);
      \draw[->,mygray] (1) -- (13);
      \draw[->,mygray](13) -- (123);
      \draw[->,mygray] (2) -- (23);
      \draw[->,mygray](23) -- (123);
      \draw[->,mygray] (3) -- (13);
      \draw[->,mygray] (3) -- (23);
    \end{tikzpicture}
  }
  }
  \caption{A possible ordering of the classes of \figurename~\ref{fig:tripartite-app} is $1, 4, 3, 2$.}
  \label{fig:hasse}
\end{figure}

We first use the inclusion relation on the power set of $\{1,\ldots,K\}$
to order the type sets $\types_i$ for $i \in \classes$.
Specifically,
we consider a topological ordering of these sets
induced by their Hasse diagram,
so that a given type set is not a subset of any type set with a lower index.
An example is shown in \figurename~\ref{fig:typesets}.
The tokens of a class with a given type set can thus overtake (in the first queue)
the tokens of all classes with a lower type set index.
Only classes with the same type set are not
dissociated.

Symmetrically, we use the inclusion relation on the power set of $\{1,\ldots,S\}$
to order the server sets $\servers_i$ for $i \in \classes$.
We consider a topological ordering of these sets
induced by their Hasse diagram,
so that a given server set is not a subset of any server set with a lower index,
as illustrated in \figurename~\ref{fig:serversets}.
The tokens of a class with a given server set can thus overtake (in the second queue)
the tokens of all classes with a lower server set index.
Thanks to
the separability assumption,
if two classes are not dissociated by their type sets,
then they are dissociated by their server sets.

This allows us to define a permutation of the classes as follows:
first, we order classes by increasing type set order,
and then, we order the classes that have the same type set by increasing server set order.
The separability assumption ensures that all classes are eventually sorted.
The tokens of a given class can overtake the tokens of all classes with a lower index,
either in the queue of available tokens or in the queue of tokens held by jobs in service (or both).

\paragraph{Moving class-$N$ tokens}

Using the two operations
\emph{circular shift}
and \emph{overtaking},
we show that,
from any given state,
we can reach the state where all class-$N$ tokens are gathered at
some selected position in one of the two queues,
while the position of the other tokens is unchanged.
We proceed by moving class-$N$ tokens one after the other,
starting with the token that is closest to the destination (in number of tokens to overtake)
and finishing with the one that is furthest.

Consider the class-$N$ token that is closest to the destination but not well positioned yet (if any).
This token can move to the destination by overtaking its predecessors one after the other.
Indeed, the token that precedes our class-$N$ token has a class between $1$ and $N-1$,
so that our class-$N$ token can overtake it in (at least) one of the two queues.
By applying many circular shifts if necessary,
we can reach the state where this overtaking can occur.
Once this state is reached,
our class-$N$ token
can then overtake its predecessor,
therefore arriving one step closer to the destination.
We reiterate this operation until our class-$N$ token is well positioned.

For example,
consider the state of \figurename~\ref{fig:initial}
and assume that we want to move all tokens of class $2$
between the
two tokens of classes $1$ and $3$ that are closest to each other.
One of the class-$2$ tokens is already in the correct position.
Let us consider the next class-$2$ token,
initially positioned between tokens of classes $3$ and $4$.
We first apply circular shifts
to reach the state depicted in \figurename~\ref{fig:final}.
In this state, there is a nonzero probability that our class-$2$ token
overtakes the class-$3$ token,
which would bring our class-$2$ token directly in the correct position.

\paragraph{Proof by induction}

We finally prove the stated irreducibility result
by induction on the number $N$ of classes.
For $N = 1$,
applying circular shifts is enough to show the irreducibility
because all tokens are exchangeable.
We now give the induction step.

Let $N > 1$.
Assume that
the Markov process defined by the state of any tandem network with $N-1$ classes
that satisfies the positive service rate and separability assumptions is irreducible.
Now consider a tandem network with $N$ classes that also satisfies these assumptions.
We have shown that,
starting from any feasible state,
we can move class-$N$ tokens at a position where they do not prevent other tokens from overtaking each other.
In particular,
to reach a state from another one,
we can first focus on ordering the tokens of classes $1$ and $N-1$, as if class-$N$ tokens were absent.
This is equivalent to ordering tokens in a tandem network with $N-1$ classes that satisfies the positive service rate and separability assumptions.
This reordering is feasible by the induction assumption.
Once it is performed,
we can move class-$N$ tokens in a correct position,
by applying the same type of transitions as in the previous paragraph.

\end{document}